\documentclass[byrevtex,prb,floatfix,a4paper,twocolumn]{revtex4}
\usepackage{amssymb,amsmath,bm}
\usepackage{graphics}
\newcommand*{\bs}{ \ensuremath{\bm{\sigma\!}}}
\newcommand*{\btau}{\ensuremath{\bm{\tau}}}

\newcommand*{\br}{\ensuremath{\bm{r}}}

\newcommand*{\bQ}{\ensuremath{\bm{Q}}}
\newcommand*{\bfG}{\ensuremath{\bm{G}}}

\newcommand*{\bW}{\ensuremath{\bm{W}}}

\newcommand*{\bk}{\ensuremath{\bm{\kappa}}}
\newcommand*{\avel}{\bigl\langle}
\newcommand*{\aver}{\bigr\rangle}

\newcommand*{\average}[1]{\avel #1 \aver}

\providecommand{\abs}[1]{\lvert#1\rvert}
\newcommand*{\likeatop}[2]{\genfrac{}{}{0pt}{1}{#1}{#2}}

\begin{document}

\title{\bf {Universal consequences of the presence of excluded volume interactions in
dilute polymer solutions undergoing shear flow }}
\author{K. Satheesh Kumar}
\altaffiliation[Present address: ]{Department of Physics, POSTECH
 San 31, Hyoja-dong, Nam-gu, Pohang, Kyungbuk, 790-784
 Republic of Korea}
\author{J. Ravi Prakash}
\affiliation{ Department~of~Chemical~Engineering,
Monash~University, Clayton, Victoria\textendash 3168, Australia}
\email{ravi.jagadeeshan@eng.monash.edu.au}
\date{\today}

\begin{abstract}
The role of solvent quality in determining the universal
material properties of dilute polymer solutions undergoing
steady simple shear flow is examined. A bead-spring chain
representation of the polymer molecule is used, and the
influence of solvent molecules on polymer conformations is
modelled by a narrow Gaussian excluded volume potential
that acts pair-wise between the beads of the chain.
Brownian dynamics simulations data, acquired for chains of
finite length, and extrapolated to the limit of infinite
chain length, are shown to be model independent. This
feature of the narrow Gaussian potential, which leads to
results identical to a $\delta$-function repulsive
potential, enables the prediction of both universal
crossover scaling functions and asymptotic behavior in the
excluded volume limit. Universal viscometric functions,
obtained by this procedure, are found to exhibit increased
shear thinning with increasing solvent quality. In the
excluded volume limit, they are found to obey power law
scaling with the characteristic shear rate $\beta$, in
close agreement with previously obtained renormalization
group results. The presence of excluded volume interactions
is also shown to lead to a weakening of the alignment of
the polymer chain with the flow direction.

\end{abstract}

\maketitle

\section{Introduction \label{intro}}

The nature of the thermodynamic interactions between the
monomers in a polymer chain and the solvent molecules in
their neighborhood, determines the ensemble of spatial
configurations adopted by the chain, and as a result, has a
significant influence on all conformation dependent
properties of polymer solutions. Extensive investigations
of static polymer solutions has established that both the
temperature of the solution and the molecular weight of the
dissolved polymer, control the strength of these
interactions. Indeed their influence can be combined
together in terms of a single variable, the so-called
\textit{solvent quality} parameter, $z = v_0(1-T_\theta/T)
\sqrt{M}$, where, $v_0$ is a polymer-solvent system
dependent constant, $T$ is the temperature of the solution,
$T_\theta$ is the theta temperature, and $M$ is the
molecular weight. At large values of $z$, usually termed
the \textit{excluded volume limit}, various properties of
polymer solutions have been shown to obey power law
scalings with molecular weight. On the other hand, at
intermediate values of $z$ (the \textit{crossover} region),
the behavior of a host of different polymer-solvent systems
has been found to be collapsible onto universal master
plots with a suitable choice of the constant
$v_0$.~\cite{miyfuj81,vidron85,haygra99,beretal99}

The entire range of observed behavior has been successfully
modelled by static theories of polymer solutions by
mimicking the influence of the solvent molecules on the
conformations of a polymer coil, with the help of a
potential that acts pair-wise between the various parts of
a polymer chain. Typically, by representing the linear
polymer molecule by a bead-spring chain consisting of $N$
beads connected together by $(N-1)$ Hookean springs, more
or less repulsion by the solvent molecules is accommodated
by varying the magnitude of the pair-wise \textit{excluded
volume} interaction parameter, denoted here by $z^*$,
between the beads.~\cite{doiedw,freed,clojan,schafer} In
particular, by noting that $z^* \sim (1-{T_\theta}/{T})$,
and $N \sim M$, a quantitative prediction of both the
excluded volume limit and the crossover regime behavior of
a number of equilibrium properties has been achieved by
analytical theories, with the help of the mapping $z=z^*
\sqrt N$.

The growing recognition that polymer-solvent interactions also
play a key role in determining the behavior of polymer solutions
far from equilibrium is responsible for the recent appearance of a
number of studies aimed at obtaining a better understanding of
this
phenomenon.~\cite{ott89c,zylott91,andetal98,ciftor99,praott99,
lilar00a,pierryck00,jenetal02,pra01a,pra01b,pra02,prapra02,prapra04,ciftor04}
Due to the analytically intractable nature of the non-equilibrium
theory, a majority of these investigations have been largely
numerical in nature. As in the case of static theories,
polymer-solvent interactions have been modelled with the help of
an excluded volume potential, such as the Lennard-Jones potential,
acting pair-wise between the different parts of the polymer chain.
While a number of qualitative conclusions on the influence of
excluded volume interactions have been drawn by these studies, a
methodical program aimed at cataloging the universal behavior of
polymer solutions in a variety of flow fields in terms of the
solvent quality parameter $z$\textemdash as has been done at
equilibrium\textemdash has not yet been attempted. Nevertheless,
some progress in this direction has been made in the framework of
approximate analytical theories. In the excluded volume limit,
\"Ottinger and co-workers~\cite{ott89c,ott90,zylott91} have
examined, with the help of approximate renormalization group
methods, the universal behavior of polymer solutions undergoing
shear flow. The solvent quality crossover scaling behavior of
viscometric functions was, however, not reported by these authors.
Recently, Prakash~\cite{pra01a,pra02} has obtained, with the help
of a Gaussian approximation, universal material functions in shear
flow, both in the crossover regime and in the excluded volume
limit.

A systematic methodology by which it is possible to go beyond
approximate results, and obtain exact results (albeit within
numerical error bars) of the influence of solvent quality, in
terms of the parameter $z$, on both equilibrium  and
non-equilibrium properties, has been newly introduced by Kumar and
Prakash.~\cite{kumpra03} Basically, they showed that model
independent predictions can be obtained by extrapolating
\textit{exact}~\footnote{Italics are used here throughout to
qualify the term ``exact", with a view to indicating the fact that
results are exact only to within the numerical error bars of the
simulations. These errors can, of course, be made arbitrarily
small by suitably increasing the number of simulation
trajectories.} Brownian dynamics simulations data, acquired for
chains of finite length, to the limit of infinite chain length.
While their methodology is applicable to both regimes of behavior,
they restricted their attention to the prediction of equilibrium
and linear viscoelastic properties. In this paper, we apply their
procedure and obtain, for the first time, \textit{exact}
predictions of the universal crossover dependence of viscometric
functions on $z$, at various values of the characteristic
non-dimensional shear rate, $\beta=\lambda_{\text{p}}
\dot{\gamma}$, where $\lambda_{\text{p}}$ is a characteristic
relaxation time defined by,
\begin{equation}
\lambda_{\text{p}}=\frac{[\eta ]_0 M \eta_{\text{s}}}{N_{\text{A}}
k_{\text{B}} T} \label{lambda}
\end{equation}
Here, $k_{\text{B}}$
is Boltzmann's constant, $\eta_{\text{s}}$ is the solvent
viscosity, $[\eta ]_0$ is the zero shear rate intrinsic viscosity,
and $N_{\text{A}}$ is Avogadro's number. The same methodology is
also used here to calculate the dependence of several properties
on $\beta$, in the excluded volume limit. This includes, for
instance, the power-law scaling of viscometric functions with
$\beta$ for large values of $\beta$, and the dependence of the
orientation of the polymer coil with respect to the flow
direction, on the magnitude of $\beta$. The important question of
whether the value of the critical exponent $\nu$ (which governs
the scaling of the mean size of a polymer chain with molecular
weight in the long chain limit), is unaltered in shear flow, is
also addressed here within this framework.

In addition to excluded volume interactions, it is now well
established that it is also essential to include solvent mediated
hydrodynamic interactions between different segments of a polymer
chain, in order to obtain an accurate description of the universal
dynamics of polymer solutions. While several studies (based on
bead-spring chain models) of the combined influence of excluded
volume, hydrodynamic interaction, and even finite extensibility
effects,  on rheological properties have been reported so far,
they have generally been restricted to finite
chains.~\cite{castor91,knuetal96,jenetal02,prapra04,ciftor04}
However, Zylka and \"Ottinger~\cite{zylott91} have examined the
universal consequences of the presence of both excluded volume and
hydrodynamic interaction effects on properties far from
equilibrium, with the help of approximate renormalization group
methods. Several important advances in recent years, within the
framework of Brownian dynamics simulations, have now made it
feasible to treat hydrodynamic interaction effects, free of any
simplifying approximations, in bead-spring chain models with a
large number of beads.~\cite{jenetal00,kroetal00} In this work,
however, we confine attention to excluded volume interactions
alone, for two reasons. First, we wish to examine, in isolation,
the influence of excluded volume interactions on long chain
properties, before considering the more complex possibilities that
arise in the presence of other non-linearities. The preliminary
results of Prabhakar and Prakash,~\cite{prapra04} who show that in
fact the various non-linearities appear to be decoupled from each
other, suggests that there is some value in looking at these
effects in isolation from each other. Second, and more
practically, the procedure adopted here of extrapolating finite
chain simulation data to the long chain limit becomes
significantly more computationally intensive when both excluded
volume and hydrodynamic interactions are included. The results of
the present paper are therefore not yet directly comparable with
experiments. Nevertheless, while the more ambitious task of a
complete description of polymer solution dynamics is being
simultaneously pursued, it is felt that several important
conclusions with regard to the role of solvent quality, can still
be made within the present approach.

The plan of the paper is as follows. In sec.~\ref{basic}, we state
the governing equations, define the various properties calculated
in this work, and summarize the previously introduced procedure
for obtaining \textit{exact} predictions of properties in the
crossover regime and in the excluded volume limit. In
sec.~\ref{crossover}, asymptotic results describing the dependence
of the normalized viscosity and the normalized first normal stress
difference on $\beta$ and $z$, are presented. The excluded volume
limit dependence on $\beta$, of various properties, is discussed
in sec.~\ref{evlimit}, along with a comparison with results
obtained previously with the Gaussian approximation. Finally, the
principal conclusions of this work are summarized in
sec.~\ref{end}.

\section{BEAD-SPRING CHAINS WITH EXCLUDED VOLUME INTERACTIONS \label{basic}}

\subsection{Governing equations\label{goveqn}}

Within the framework of a numerical solution strategy, the most
appropriate model for the task of describing the rheological
behavior of dilute polymer solutions, in the limit of long chains,
is a bead-spring chain model. The internal configuration of a
bead-spring chain, suspended in a Newtonian solvent undergoing
homogeneous flow, with the beads located at positions $\br_{\nu},
\, \nu =1, \ldots, N$ with respect to an arbitrarily chosen
origin, can be specified by the $N-1$ bead connector vectors,
$\bQ_k=\br_{k+1}-\br_k, \, k=1, \ldots, N-1$, connecting beads $k$
and $k+1$. For Hookean springs, and a pairwise intra-molecular
excluded volume force, the potential energy of a bead-spring chain
is given by the expression,
\begin{equation}
\phi = \frac{1 }{ 2} \, H \, \sum_{i=1}^{N-1} \, {\bQ}_i \cdot
{\bQ}_i + \frac{1 }{ 2}   \sum_{\likeatop{ \nu, \, \mu=1 }{\nu \ne
\mu }}^N \, E \left ( {\br}_{\nu} - {\br}_{\mu} \right)
\label{poteng}
\end{equation}
where, $H$ is the spring constant, and $E \left( {\br}_{\nu} -
{\br}_{\mu} \right)$ is the excluded volume potential between the
beads $\nu$ and $\mu$ of the chain. The key ingredient, in the
procedure for obtaining universal predictions in terms of the
parameter $z$, is the use of the narrow Gaussian potential to
represent excluded volume interactions,
\begin{equation}
E \left( {\br}_{\nu} - {\br}_{\mu} \right) = \left( \frac{z^* }{
{d^*}^3} \right) k_{\text{B}} T \, \exp \left\lbrace - \frac{H }{
2 k_{\text{B}} T }\, \frac{ \br_{\nu \mu}^2 }{ {d^*}^2}
\right\rbrace \label{evpot}
\end{equation}
where, $\br_{\nu \mu} = \br_{\nu}-\br_{\mu}$, is the vector
between beads $\mu$ and $\nu$, the nondimensional parameter $z^*$,
as mentioned previously, measures the strength of the excluded
volume interaction, and $d^*$ is a measure of the range of
excluded volume interaction. Prakash and coworkers have shown, in
their examination of universal linear viscoelastic and steady
shear flow properties in the context of the Gaussian
approximation,~\cite{pra01a,pra02} and in their calculation of
universal equilibrium and linear viscoelastic properties with
Brownian dynamics simulations,~\cite{kumpra03} that the usefulness
of the narrow Gaussian potential stems from the fact that, as $N
\to \infty$, (i) the parameters $z^*$ and $N$ combine together to
form the single variable, $z=z^* \sqrt{N}$, and (ii) the choice of
$d^*$ is inconsequential because the parameter $d^*$ always
appears in the theory as the re-scaled variable $d^* / \sqrt N$.
These results form the basis of our exploration, in this work, of
the universal behavior of bead-spring chains subject to shear
flow.

As will be discussed in greater detail shortly in
sec.~\ref{matfun} below, all properties of interest in the
present work can be calculated from appropriately defined
configurational averages. Unfortunately, it is currently
infeasible to solve the nonlinear diffusion equation that
governs the time evolution of the configurational
distribution function in the presence of excluded volume
interactions. However, configurational averages can be
numerically evaluated directly from ensembles of polymer
configurations. Basically, stochastic trajectories that
describe the temporal evolution of an ensemble of chain
configurations can be generated by solving the stochastic
differential equation that governs the bead connector
vectors, with the help of Brownian dynamics
simulations.~\cite{ottbk} In terms of the non-dimensional
variables,
\begin{equation}
\bQ^*_k=\frac{1}{\ell} \, \bQ_k; \, \, t^*=\frac{t}{\lambda_H}; \,
\, \phi^*=\frac{\phi}{k_BT}; \, \, \bk^*=\lambda_H\bk
\end{equation}
where, $\ell = \sqrt{k_{\text{B}}\, T /H}$ is a length scale that
is proportional to the root mean square extension of a single
spring in the Rouse model, $\lambda_H=\zeta/4H$ is a time constant
(with $\zeta$ representing the bead friction coefficient), and
$\bk (t)$ is the traceless transpose of the velocity-gradient
tensor for the homogeneous flow field (which can be a function of
time but is independent of position), the stochastic differential
equation that governs the trajectories of the bead connector
vectors in the presence of excluded volume interactions, is given
by the expression,~\citep{pra02}
\begin{equation}
  d\bQ^*_j=\left[\bk^*\cdot\bQ^*_j-\frac14
    \sum_{k=1}^{N-1}A_{jk}\frac{\partial  \phi^*}{\partial
    \bQ^*_k}\right]      dt^*\\      +\sqrt{\frac12}\sum_{\nu-1}^N\bar
    B_{j\nu}\,d\bW^*_\nu\label{goveq}
\end{equation}
Here, $\bW_\nu^*$ is  a  Wiener   process, whose $3N-$dimensional
components satisfy,
\begin{equation}
\begin{split}
  \langle  W^*_{\nu,j}(t^*)\rangle&=0\\
  \langle  W^*_{\nu,j}(t^*)W_{\mu,k}({t^*}
  ')\rangle&=\min(t^*,{t^*} ')\delta_{jk}\delta_{\nu\mu}
\end{split}
\end{equation}
for $j,k=1,2,3$ and $\nu,\mu =1,2,3,\ldots,N$. The quantity
${\overline B}_{k \nu}$ is an $(N-1) \times N$ matrix defined by,
${\overline B_{k \nu}} = \delta_{k+1, \; \nu} - \delta_{k \nu}$,
with $\delta_{k \nu}$ denoting the Kronecker delta, and $A_{jk}$
is the Rouse matrix,
\begin{equation}
A_{jk}=\sum_{\nu=1}^{N}\, {\overline B}_{j \nu} {\overline B}_{k
\nu} = \begin{cases}
 2 & \text{for $\abs{j-k} = 0 $}, \\
-1 & \text{for $\abs{j-k} =1 $},\\
 0 & \text{ otherwise}
\end{cases}
\end{equation}

Before discussing the two different simulation schemes used
here for solving Eq.~(\ref{goveq}) to obtain predictions of
the universal crossover and excluded volume limit
behaviors, it is appropriate to first introduce all the
properties, and their defining equations, that are
calculated in the present work.

\subsection{Material functions\label{matfun}}

We confine our attention in this work to steady simple
shear flows, which are specified by the following matrix
representation for the tensor $\bk$ in a laboratory-fixed
coordinate system,
\begin{equation}
\bk=\dot{\gamma} \begin{pmatrix} 0 & 1 & 0 \\
0 & 0 & 0 \\ 0 & 0 & 0 \end{pmatrix}
\end{equation}
where, $\dot{\gamma}$ is the constant shear rate.

The principle material functions of interest here are the
polymer contribution to the viscosity, $\eta_{\text{p}}$,
and the first normal stress difference, $\Psi_1$. Since
hydrodynamic interactions have not been included in the
present model, the second normal stress difference is
identically zero.~\cite{birdetal2} In terms of the
components of the polymer contribution to the stress
tensor, $\tau_{xy}^{\text{p}}, \tau_{xx}^{\text{p}}$, etc,
$\eta_{\text{p}}$, and $\Psi_1$, can be obtained from the
following relations,~\citep{birdetal1}
\begin{equation}
\tau_{xy}^{\text{p}} = - {\dot \gamma}\, \eta_{\text{p}} \,
; \quad \quad \tau_{xx}^{\text{p}}- \tau_{yy}^{\text{p}} =
- {\dot \gamma^2}\, \Psi_1 \label{sfvis}
\end{equation}
Within the framework of polymer kinetic theory, the polymer
contribution to the stress tensor, $\btau^{\text{p}}$, can
be calculated by either the \textit{Kramers} or the
\textit{Giesekus} expressions. While the latter expression
becomes invalid in the presence of hydrodynamic
interactions, it has been shown to remain valid in the
presence of excluded volume interactions.~\cite{pra01b} The
relative magnitude of the variance associated with the use
of either of these expressions in Brownian dynamics
simulations has been examined previously by \citet{pra02}
and it has been shown that the Giesekus expression always
leads, particularly at low shear rates, to a smaller value
of variance.  For this reason, we have used the Giesekus
expression in our work, which at steady state has the form,
\begin{equation}
\btau^\text{p} = -  \frac{n_{\text{p}} \, \zeta}{2}
\sum_{m, n =1}^{N-1} C_{m n} \left\lbrace \, \bk \cdot
\average{\bQ_m \bQ_n}  + \average{\bQ_m \bQ_n} \cdot \bk^T
\right\rbrace \label{gstress}
\end{equation}
where, $C_{m n}$ is the Kramers matrix. The Kramers matrix
is the inverse of the Rouse matrix, and is defined by,
\begin{equation}
C_{m n} = \sum_{\nu=1}^{N}\, B_{\nu m} B_{\nu n} = \min \,
(m,n) - \frac{m \, n}{ N} \label{kmatrix}
\end{equation}
Here, the $N \times (N-1)$ matrix $B_{\nu k}$ is defined by,
$B_{\nu k} = {k}/{N} - \Theta \, (k-\nu)$, with $\Theta \,
(k-\nu)$ denoting a Heaviside step function. As will be discussed
in greater detail in sec.~\ref{simscheme} below, we further reduce
the variance in our simulations by implementing a variance
reduction procedure based on the method of control
variates.~\cite{ottbk}

In the limit of zero shear rate, the ratio $U_{\Psi\eta}$,
which is defined by the expression,
\begin{equation} \label{UPE}
U_{\Psi\eta}=\frac{n_p \, k_BT\Psi_{1}}{\eta^2_{p}}
\end{equation}
where, $n_p$ is the number density of polymers, has been shown to
have a universal value, both in theta solutions,~\cite{ottbk} and
in good solvents.~\cite{kumpra03} In the context of a dumbbell
model, Prakash and \"Ottinger~\cite{praott99} have evaluated its
dependence on the non-dimensional shear rate $\lambda_H \dot
\gamma$. In this work, we seek to examine the universal dependence
of $U_{\Psi\eta}$ on the characteristic shear rate $\beta$, in the
excluded volume limit.

In addition to rheological properties, it is also of
interest to examine the \textit{anisotropy} induced in any
arbitrary tensorial property of the solution, due to the
orientation and deformation of polymer coils caused by the
flow field. The precise definition of anisotropy is taken
up shortly below. Of particular relevance in this work, is
the change in the degree of anisotropy brought about by the
presence of excluded volume interactions.

In order to obtain an estimate of the anisotropy of any
tensorial quantity $\bs$, we begin by noting that in simple
shear flows $\bs$ must reflect the symmetry of the flow
field. In particular, since shear flows are invariant when
the direction of the \textit{z}-axis is
reversed,~\citep{birdetal1} the \textit{xz}, \textit{yz},
\textit{zx} and \textit{zy} components of $\bs$ must be
zero. As a result, $\bs$ must have the following matrix
representation in the laboratory-fixed coordinate system,
\begin{equation}
\bs=\begin{pmatrix} \sigma_{xx} & \sigma_{xy} & 0 \\[4pt]
\sigma_{yx} & \sigma_{yy} & 0 \\[4pt] 0 & 0 & \sigma_{zz}
\end{pmatrix}
\label{bs}
\end{equation}
It is clear from Eq.~(\ref{bs}) that the \textit{z}-axis is
a principal direction. It follows that the other two
principal directions must lie in the \textit{xy} plane. We
denote the principal axes in this plane by $\bar x$ and
$\bar y$, with the $\bar x$ axis making an angle
$\chi_\sigma$ with the \textit{x}-axis. Thus by rotating
the coordinate system by $\chi_\sigma$, the matrix that
represents $\bs$ can be diagonalized. By exploiting the
rules for change of tensor components, the orientation of
the $(x,y,z)$ coordinate system relative to the $(\bar x,
\bar y, \bar z)$ coordinate system can be calculated to be,
\begin{equation}
\tan 2 \chi_\sigma =\frac{2 \sigma_{xy}}{\sigma_{xx} -
\sigma_{yy}} \label{chi}
\end{equation}
With the help of symmetry arguments, one can show that the
orientation angle $\chi_\sigma = \pi/4$ close to equilibrium. As
the shear rate increases, $\chi_\sigma$ is expected to decrease to
zero, reflecting the increasing alignment of the polymer coils
with the flow direction. The orientation angle $\chi_\sigma$,
therefore, serves as a measure of the anisotropy of
$\bs$.~\cite{pierryck00}

It is straightforward to show that the orientation angle
associated with the polymer contribution to the stress tensor,
$\chi_\tau$, is given by,~\cite{bossott95,pierryck00}
\begin{equation}
\tan 2 \chi_\tau =\frac{2 \tau_{xy}^{\text{p}}
}{\tau_{xx}^{\text{p}} - \tau_{yy}^{\text{p}}
}=\frac{m_\tau}{\beta} \label{chitau}
\end{equation}
where the quantity $m_\tau$, defined by the above expression, is
frequently referred to as the \textit{orientation
resistance}.~\cite{bossott95} In the Rouse model, $m_\tau$ is a
constant, independent of the shear rate, with a value 2.5. In a
similar manner, the orientation angle associated with the radius
of gyration tensor, $\bfG$, is given by the
expression,~\cite{bossott95,pierryck00}
\begin{equation}
\tan 2 \chi_G =\frac{2 G_{xy}}{G_{xx} - G_{yy}}=\frac{m_G}{\beta}
\label{chiG}
\end{equation}
where, $m_G$ is the corresponding orientation resistance, and
$G_{xy}$ etc., represent components of the tensor $\bfG$, defined
by,
\begin{equation}
\bfG = \frac{1}{N} \sum_{\nu=1}^{N} \avel (\br_{\nu} -\br_c)
(\br_{\nu} -\br_c)\aver \label{radg}
\end{equation}
Here, $\br_c$ denotes the position of the center of mass. In the
Rouse model, $m_G = 1.75$. It is worthwhile noting that while the
orientation of the macromolecule measured by flow birefringence
coincides with $\chi_\tau$ (according to the ``stress optical"
law), $\chi_G$ is the orientation angle measured by static light
scattering.~\cite{bossott95} In this work, we show that in the
presence of excluded volume interactions, both $m_\tau$ and $m_G$
are not constant, but rather are functions of the characteristic
shear rate $\beta$, and we examine this dependence in the excluded
volume limit.

\subsection{Simulation schemes \label{simscheme}}

Two different simulation schemes have been introduced and
discussed in detail by Prakash and co-workers in order to
find the universal properties of polymer solutions in the
crossover region, and in the excluded volume limit,
respectively.~\cite{pra02,kumpra03} Here, we summarize them
briefly.

The universal dependence of various properties on the
parameter $z$ in the long chain limit is found by a two
step procedure: (i) Simulations are carried out for
increasing values of $N$, keeping the value of $z$ ($=z^*
\sqrt{N}$), constant, and (ii) The accumulated finite chain
data for any property is then extrapolated to the limit $N
\to \infty$. The crossover behavior of the property is
obtained  by repeating steps (i) and (ii) for a number of
values of $z$. In the Gaussian approximation, the value of
the parameter $d^*$ was held constant during the process of
accumulating finite chain data,~\cite{pra01a,pra02} and it
was shown that the particular choice of value for $d^*$ was
inconsequential in the limit $N \to \infty$. On the other
hand, while obtaining the crossover behavior of equilibrium
and linear viscoelastic properties with Brownian dynamics
simulations,~\cite{kumpra03} the parameter $d^*$ was chosen
such that,
\begin{equation}
d^*=k\, ({z^*})^{{1/5}} \label{krelation}
\end{equation}
where, $k$ is an arbitrary constant. Note that, for a fixed
value of $z$, since $z^* \to 0$ (or equivalently, $T \to
T_\theta$), as $N \to \infty$, this choice of $d^*$ implies
that the asymptotic limit is reached along trajectories in
the ($z^*, d^*$) parameter space that converge to the
origin. It was shown by Kumar and Prakash~\cite{kumpra03}
that choosing $d^*$ values according to
Eq.~(\ref{krelation}), permits the use of larger step sizes
in the numerical integration scheme. The independence of
the asymptotic results from the particular choice of $k$
(and consequently the trajectory in the ($z^*, d^*$)
parameter space), was also established in
ref.~\onlinecite{kumpra03}. Since Brownian dynamics
simulations are used in the present work to obtain the
crossover behavior, $d^*$ values will be chosen according
to Eq.~(\ref{krelation}).

Universal properties in the excluded volume limit ($z \to
\infty$) are obtained by accumulating finite chain data for
constant values of $z^*$ (which is equivalent to the
temperature $T$ being constant), followed by extrapolation
to the infinite chain length limit.  Since for any constant
value of $z^*$, $z \to \infty$, as $N \to \infty$, we
expect that properties obtained in the excluded volume
limit will be independent of the particular choice of
$z^*$. Prakash and co-workers~\cite{pra02,kumpra03} have
shown that this is indeed the case, for the situations
examined previously by them. As will be discussed in
greater detail subsequently, this independence will be used
in the present work to indicate the adequacy or otherwise
of the accumulated finite chain data.

The various configurational averages required to calculate
the properties listed in sec.~\ref{matfun} above were
obtained with a second order predictor-corrector Brownian
dynamics simulation algorithm originally proposed by
Iniesta and Garcia~de~la~Torre.~\cite{iniesta90} A detailed
discussion of the algorithm, adapted for the problem at
hand, can be found in ref.~\onlinecite{pra02}. In a typical
simulation, roughly 80,000 trajectories, with Gaussian
initial distributions, were generated, and the integration
was carried out until a stationary state was attained.
Message Passing Interface (MPI) clusters were used to
distribute the trajectories across several processors.
Since the present algorithm belongs to the class of
embarrassingly parallel algorithms, the wall time scales
with the number of processors used. Typically, three
different time steps $\Delta t^* = 1.0, \, 0.6, \text{and}
\, 0.5$ were used, and the results were then extrapolated
to zero time step using the subroutine \texttt{TEXTRA}
suggested by \"Ottinger~\cite{ottbk}. This extrapolation
procedure enabled the simulations to be carried out at
relatively large time steps.

A key ingredient in the simulation strategy adopted here,
as mentioned perviously, is the use of a variance reduction
procedure based on the method of control variates, in order
to reduce the statistical error in the simulations. This is
crucial because the magnitude of the error in the finite
chain data significantly affects the accuracy of the
extrapolation to the infinite chain length limit. Since a
fairly involved discussion of the procedure has been given
previously in ref.~\onlinecite{kumpra03}, it is not
repeated here. However, a few salient features are
highlighted below.

The essence of the scheme (and the vital factor that decides its
success or otherwise) is to find a control variable whose mean
value is known exactly, and whose fluctuations are correlated with
the variable whose variance we are interested in reducing. It was
found by Kumar and Prakash,~\cite{kumpra03} that the Rouse model
serves as an excellent source of control variables for all the
equilibrium and linear viscoelastic properties that were evaluated
by them. In this work, we have again used the Rouse model as the
source of control variables, and found that as before, this leads
to a significant reduction in the variance of all properties. At
first sight, this result is somewhat surprising. In the case of
equilibrium and linear viscoelastic properties, since only
equilibrium averages were evaluated, both the stochastic
differential equation, Eq.~\ref{goveq}, and the stochastic
differential equation in the Rouse model, were solved with the
term involving $\bk$ set equal to zero. In the present instance,
since we are interested in properties at finite shear rates, both
equations were integrated with the $\bk$ term in place. In spite
of including this term in the Rouse model, as is well known,
viscometric functions predicted by it are constant and independent
of shear rate. This is not the case, however, in the presence of
excluded volume interactions, where, as we shall see shortly,
viscometric functions decrease with an increase in shear rate. The
success in the variance reduction strategy suggests, therefore,
that even though the Rouse viscometric functions are independent
of shear rate, the fluctuations in the Rouse control variables
remain correlated with their corresponding variables in the
presence of excluded volume interactions. The reason for this
occurance might lie in the fact that with increasing shear rate,
the beads of the chain are drawn further apart, and the resultant
weakening of the excluded volume interactions actually brings the
results of the two models closer together.

\section{UNIVERSAL PROPERTIES \label{universal}}

\subsection{Crossover behavior\label{crossover}}

In this subsection, we are concerned with obtaining the crossover
behavior of the universal ratios, ${\eta_p}/{\eta_{p,0}}$, and
$\Psi_1/\Psi_{1,0}$ (where, $\eta_{p,0}$ and $\Psi_{1,0}$ are the
zero shear rate polymer contributions to the viscosity and first
normal stress difference, respectively), as a function of the
characteristic shear rate $\beta$, using the simulation scheme
described in sec.~\ref{simscheme} above. A strict pursuance of
this procedure would require the construction of each of these
ratios at various values of $N$, using equilibrium simulations (as
described previously in ref.~\onlinecite{kumpra03}) to find
$\eta_{p,0}$ and $\Psi_{1,0}$, and simulations at finite shear
rates (as described above) to find $\eta_{p}$ and $\Psi_{1}$,
followed by extrapolation to the long chain limit. Unfortunately,
this procedure suffers from a serious problem. Since both the
denominator and the numerator are found by Brownian dynamics
simulations, the ratios have relatively large statistical error
bars, which makes their extrapolation to the $N \to \infty$ limit
highly inaccurate. Fortunately, this difficulty can be overcome by
adopting the following procedure.

We first note that,
\begin{align}
  \lim_{N\to\infty}
  \frac{\eta_p}{\eta_{p,0}} & =\frac{\lim_{N\to\infty}(\eta_p/\eta_p^R)}
  {\lim_{N\to\infty}{(\eta_{p,0}/\eta_{p}^R})} \\
 \lim_{N\to\infty}
  \frac{\Psi_1}{\Psi_{1,0}} & =\frac{\lim_{N
  \to\infty}(\Psi_{1}/\Psi_1^R)}
  {\lim_{N\to\infty}{(\Psi_{1,0}/\Psi_1^R)}}
\end{align}
where, $\eta_p^R$ and $\Psi_{1}^R$ are the viscosity and first
normal stress difference predicted by the Rouse model,
respectively,~\cite{birdetal2}
\begin{align}
\eta_p^R&=n_p\lambda_Hk_BT\left[\frac{(N^2-1)}{3}\right]\label{rouseeta}\\
\Psi_{1}^R &= 2 n_p \, \lambda_H^2k_BT
\left[\frac{(N^2-1)(2N^2+7)}{45}\right]\label{rousepsi}
\end{align}
Kumar and Prakash have previously evaluated the ratios
$\eta_{p,0}/\eta_{p}^R$ and $\Psi_{1,0}/\Psi_1^R$ in the long
chain limit.~\cite{kumpra03} For the purpose of evaluating
${\eta_p}/{\eta_{p,0}}$ in the long chain limit, therefore, it
suffices to evaluate the ratios, $\eta_{p}/\eta_{p}^R$ and
$\Psi_{1}/\Psi_1^R$, as $N \to \infty$. The simulation scheme
described in sec.~\ref{simscheme} above has, consequently, been
used here to evaluate these ratios at various values of $z$.

\begin{figure}
  \resizebox{8.5cm}{!}{\includegraphics*{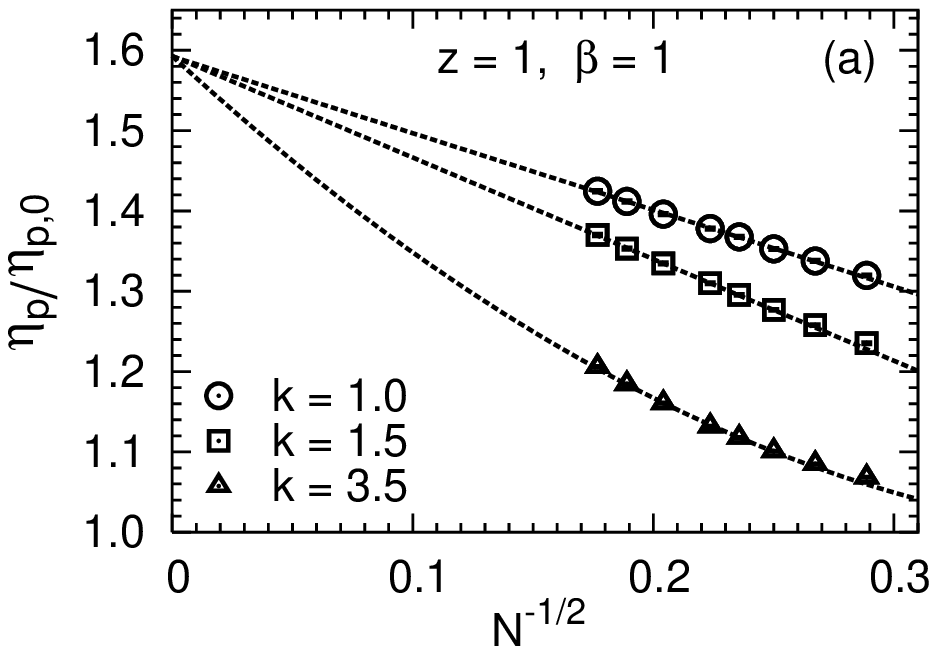} }
\resizebox{8.5cm}{!}{\includegraphics*{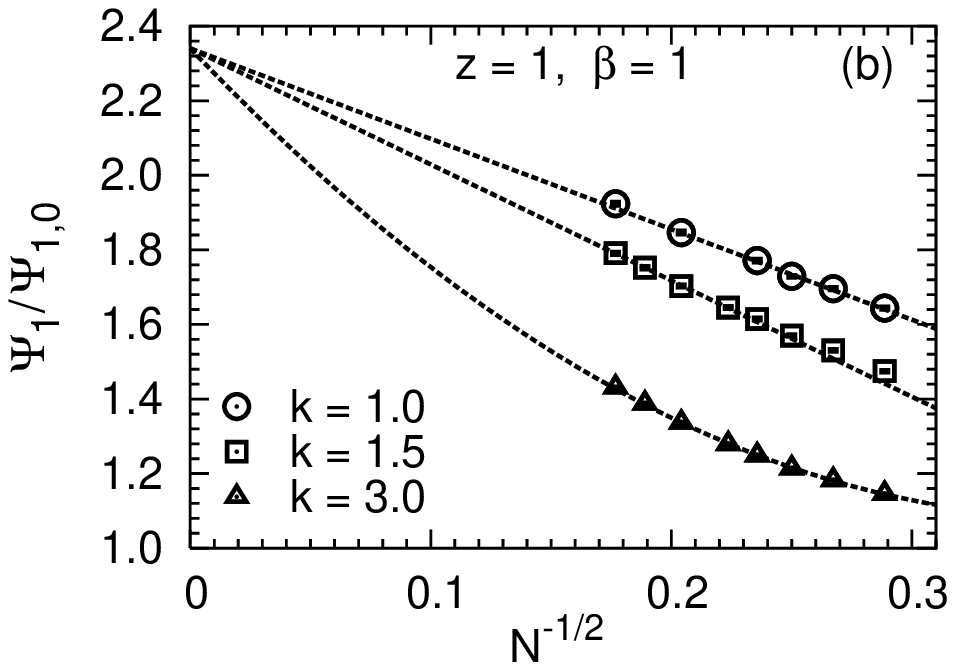} }
  \vskip5pt
  \caption{(a) The ratio of the viscosity to the Rouse model viscosity,
  and (b) The ratio of the first normal stress difference to
    the Rouse model first normal stress difference, versus
    $1/\sqrt{N}$ at $z=1$ and $\beta=1$, for three different values of
    $k$. The symbols, with error bars, represent the
    simulation results, while the lines are least-square curve
    fits to the data.
  \label{etaextra}}
\end{figure}

Figures~\ref{etaextra} clearly indicate the independence of the
extrapolated values of $\eta_{p}/\eta_{p}^R$ and
$\Psi_{1}/\Psi_1^R$, in the $N \to \infty$ limit, from the
trajectories in the ($z^*, d^*$) parameter space used to obtain
them. The parameters $z$ and $\beta$ have been held constant at a
value of unity, while carrying out simulations at increasing
values of $N$, for three different values of $k$. In each
simulation, $z^*=z/\sqrt{N}$, while $d^*$ is found from
eq~\ref{krelation}. In order to keep $\beta$ constant in each
simulation, the non-dimensional shear rate $\lambda_H \, \dot
\gamma$ used in the simulation, at each value of $N$, has been
calculated from the expression, $\lambda_H \, \dot \gamma = \beta
\, (\lambda_H n_p k_BT / \eta_{p,0})$, which can be derived from
Eq.~\ref{lambda} for a dilute polymer solution. As mentioned
earlier, ($\eta_{p,0} / \lambda_H n_p k_BT $) can be evaluated by
carrying out equilibrium simulations for the same set of parameter
values as those used in the finite shear rate simulations.
Extrapolation of the finite chain data has been carried out by
plotting simulation results versus $1/\sqrt{N}$, since
Prakash~\cite{pra01b} has shown previously that leading order
corrections to the infinite chain length limit, of various
material properties, are of order $1 /\sqrt N$. Furthermore, as
mentioned earlier, the parameter $d^*$ always occurs in the theory
as the ratio $d^* / \sqrt N$. The asymptotic values obtained in
this manner correspond to the viscosity and first normal stress
difference ratios for a $\delta$-function excluded volume
potential, at $z=1$ and $\beta=1$. The same procedure was then
repeated for various values of $z$ and $\beta$.

\begin{figure}
 \resizebox{8.5cm}{!}{\includegraphics*{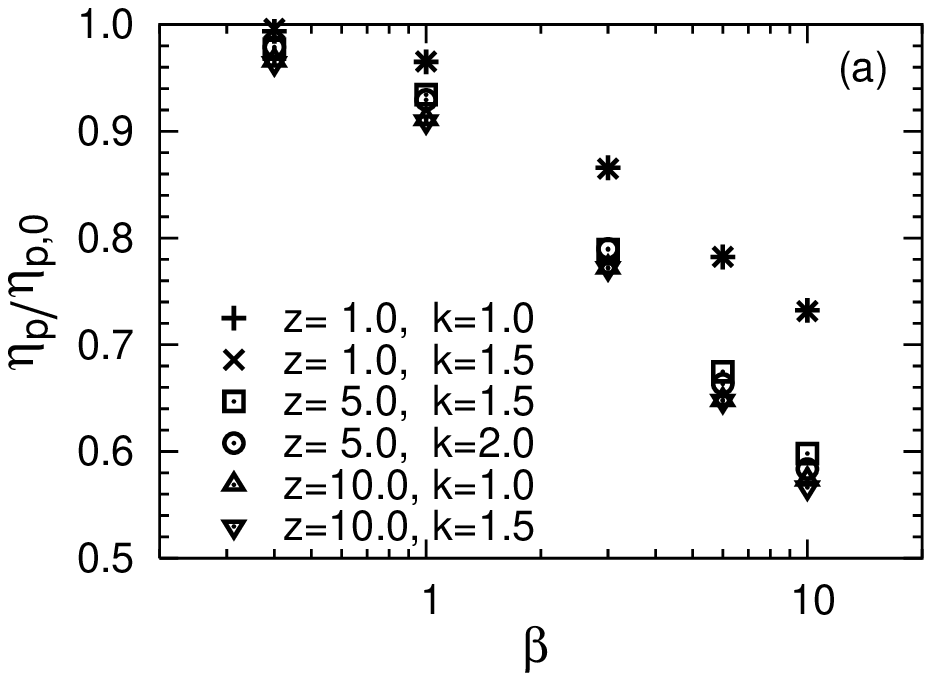} }
  \resizebox{8.5cm}{!}{\includegraphics*{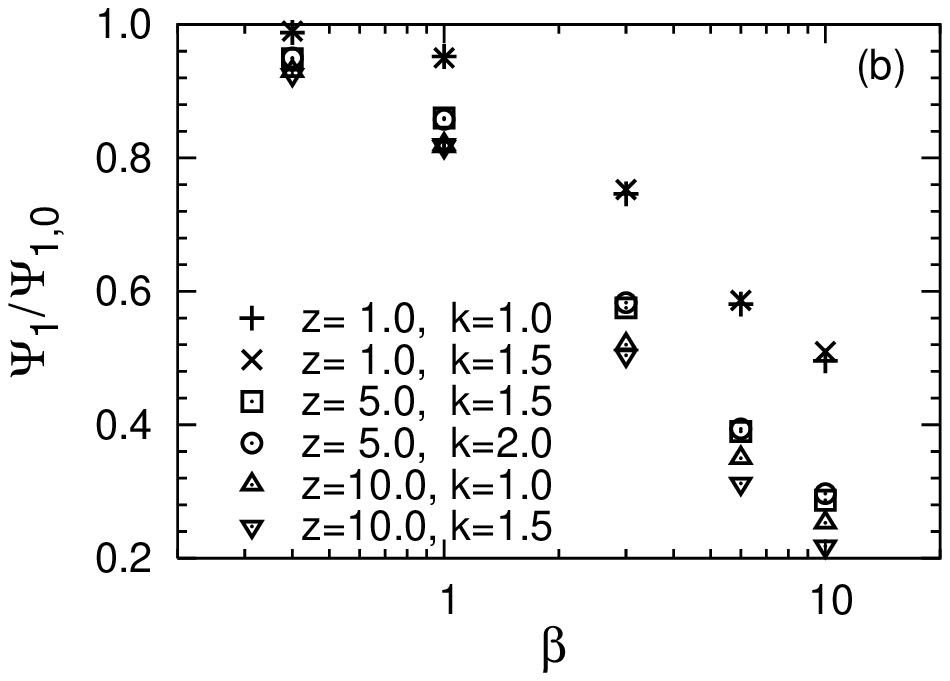} }
  \caption{(a) Asymptotic nondimensional viscosity ratio
    $(\eta_p/\eta_{p,0})$, and (b) Asymptotic nondimensional first normal
    stress difference ratio $(\Psi_1/\Psi_{1,0})$, versus the characteristic shear rate
    $\beta$,
    for various values of $z$ and $k$
\label{psiextra}}
\end{figure}

The universal dependence of ${\eta_p}/{\eta_{p,0}}$, and
$\Psi_1/\Psi_{1,0}$ on $\beta$, at various values of $z$, obtained
by combining the infinite chain length results for the ratios
$\eta_{p}/\eta_{p}^R$ and $\Psi_{1}/\Psi_1^R$, with the infinite
chain length zero shear rate ratios evaluated previously by Kumar
and Prakash,~\cite{kumpra03} are presented in
Figs.~\ref{psiextra}. At each value of $z$, extrapolated data for
two different values of $k$ are displayed in order to clearly
delineate the limit of $\beta$ up to which the current asymptotic
results are valid. In essence, beyond some threshold value of the
characteristic shear rate, say $\beta^\dag$, the lack of
coincidence of data for two different values of $k$, implies that
the present finite chain data, accumulated for chains with $N \le
36$, is  no longer sufficient to obtain an accurate extrapolation.
For $\beta > \beta^\dag$, the flow field begins to probe model
dependent length scales, and scale invariance, which is
responsible for the observed model independence, no longer exists.
The results in Figs.~\ref{psiextra} suggest that $\beta^\dag$ for
$\Psi_1/\Psi_{1,0}$ is smaller than that for
${\eta_p}/{\eta_{p,0}}$, and in both cases, it decreases with
increasing $z$.

\begin{figure}
  \begin{center}
    \resizebox{8.5cm}{!}{\includegraphics*{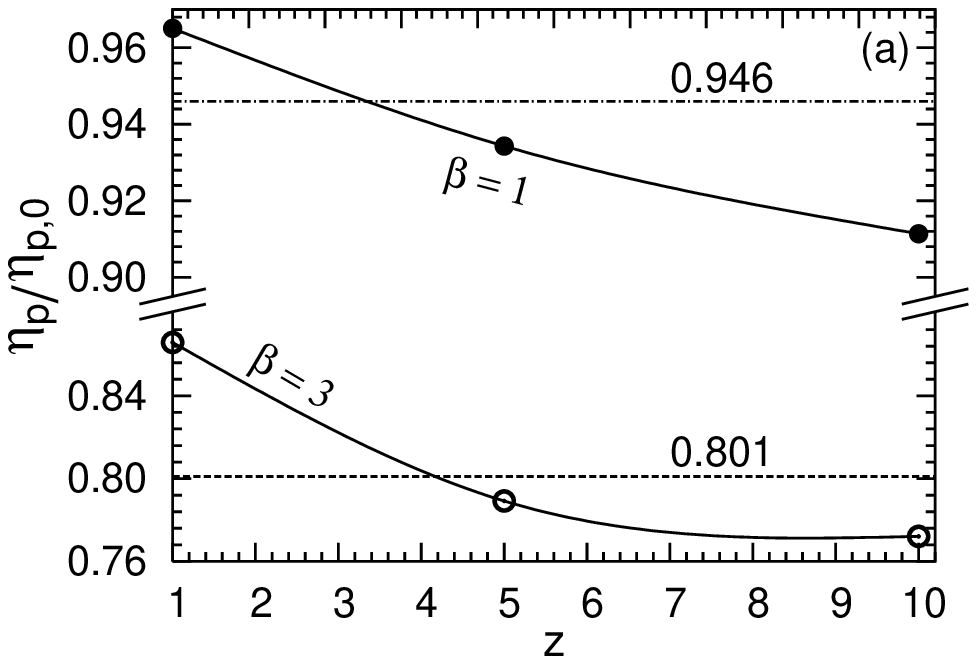} }
    \resizebox{8.5cm}{!}{\includegraphics*{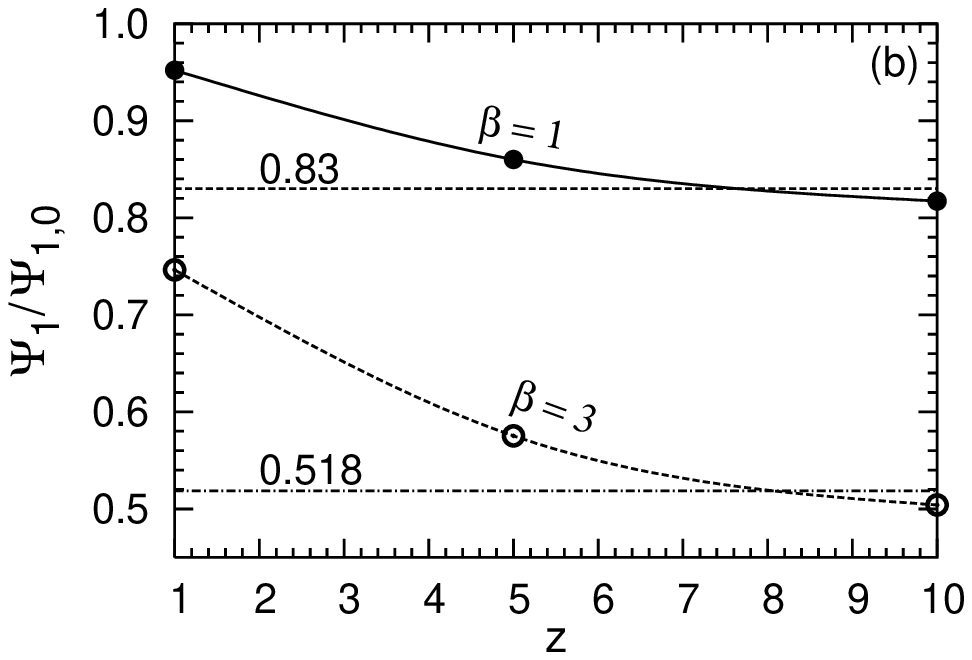} }
  \end{center}
  \caption{(a) Asymptotic non-dimensional viscosity ratio
    ($\eta_p/\eta_{p,0}$), and (b) Asymptotic non-dimensional first normal
    stress difference ratio $(\Psi_1/\Psi_{1,0})$,
    plotted against $z$ for two different values
    of $\beta$. Symbols denote values obtained in the crossover
    regime, while the horizontal lines correspond to values of
    the ratios in the excluded volume limit.}
  \label{crossev}
\end{figure}

It is clear from Figs.~\ref{psiextra} that the presence of
excluded volume interactions leads to shear thinning, and that the
extent of shear thinning increases with increasing solvent
quality. However, the incremental increase in shear thinning with
increasing values of $z$, for a given value of $\beta$, appears to
decrease as $z$ increases. This can be seen more clearly from
Figs.~\ref{crossev}, where the asymptotic viscometric ratios are
plotted versus the solvent quality $z$, at two representative
values of $\beta$. In each case, the ratios decrease, but with
decreasing rate, as the solvent quality increases, before
appearing to level off at a final solvent quality independent
value, in the limit of large $z$. In this excluded volume limit,
consequently, universal viscometric functions would depend only on
$\beta$, and on no other parameter.

It is not possible here to explore further the approach of the
crossover regime to the excluded volume limit, because the present
finite chain data is not sufficient to obtain accurate
extrapolations. Obtaining such data is, unfortunately, currently
prohibitively expensive computationally. Nevertheless, behavior in
the excluded volume limit itself can be directly obtained, as has
been discussed earlier. The horizontal lines in
Figs.~\ref{crossev} correspond to excluded volume limit values of
the viscometric ratios obtained by these means, the details of
which will be described in greater detail shortly. It is
immediately apparent that the large $z$ asymptote of the crossover
regime does not coincide with the excluded volume limit. This is
in complete contrast to the universal results, in steady shear
flow, obtained earlier by Prakash~\cite{pra02} with the Gaussian
approximation. While the general behavior in the crossover regime
predicted by the Gaussian approximation is identical to that
described above, values of viscometric ratios for $z \gg 1$ were
found to smoothly approach those obtained in the excluded volume
limit. This perplexing behavior, where predictions of the Gaussian
approximation differ qualitatively from those of \textit{exact}
Brownian dynamics simulations, was also observed recently by Kumar
and Prakash in their attempt to predict the behavior of universal
equilibrium and linear viscoelastic ratios.~\cite{kumpra03} Before
discussing the origins of this behavior in greater detail,
however, it is necessary to complete the presentation of a few
further results obtained in the excluded volume limit.

\subsection{Excluded volume limit \label{evlimit}}

\begin{figure}
  \resizebox{8.5cm}{!}{\includegraphics*{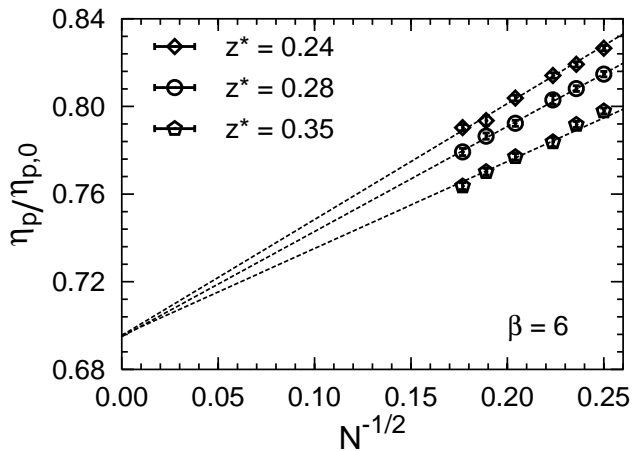} }
  \caption{Non-dimensional viscosity ratio $\eta_p/\eta_{p,0}$,
  plotted against $1/\sqrt{N}$, for three different values of
  $z^*$, and for $\beta=6$. Symbols are predictions of Brownian
  dynamics simulations, while lines are
  least square curve fits through the data. }\label{evetaextra}
\end{figure}

\begin{figure}[tb]
  \resizebox{8.5cm}{!}{\includegraphics*{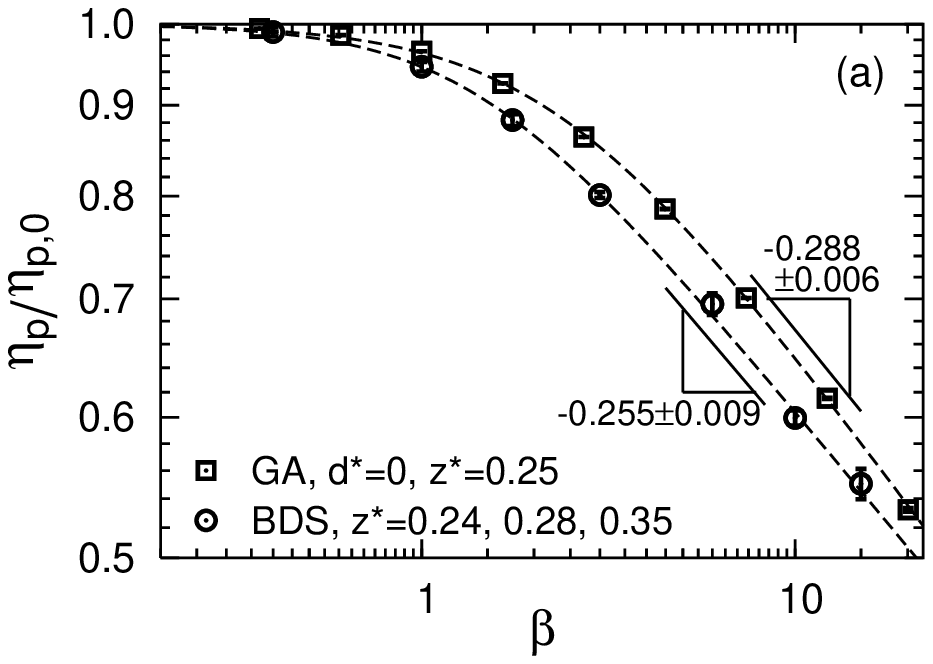} }
  \resizebox{8.5cm}{!}{\includegraphics*{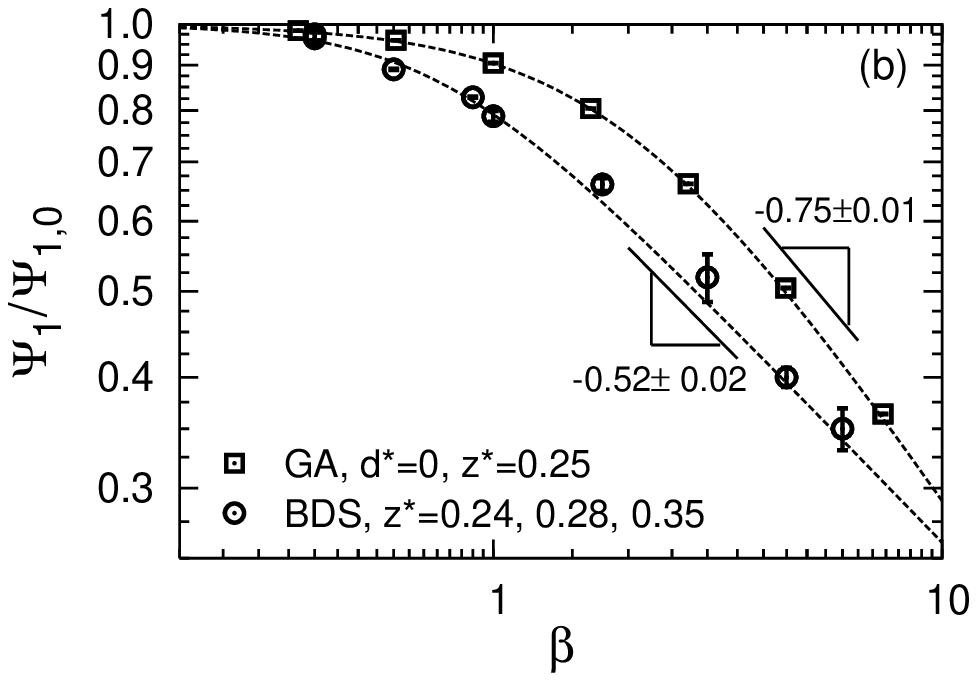} }
  \caption{(a) Asymptotic non-dimensional ratio
  $\eta_p/\eta_{p,0}$, and (b) Asymptotic
  nondimensional ratio $\Psi_1/\Psi_{1,0}$, for various values of
  $\beta$, in the excluded volume limit. Circles represent Brownian dynamics
  simulation  data, while squares represent results of the
  Gaussian approximation. The dashed lines are fits to the data
  using the Carreau-Yasuda model, Eqs.~(\ref{eqcy}). }
  \label{evpsiextra}
\end{figure}

Predictions in the excluded volume limit are obtained, as
described in sec.~\ref{simscheme} above, by extrapolating finite
chain data acquired for various constant values of $z^*$, to the
infinite chain length limit. Figure~\ref{evetaextra} is an
illustrative example, where simulations have been carried out at
several values of $N$, for $z^*=0.24, 0.28$, and $0.35$,
respectively, and at a fixed value of $\beta=6$, in order to
acquire data for the ratio $\eta_p/\eta_{p,0}$. The common
extrapolated value, in the limit $N \to \infty$, for all the three
values of $z^*$, is the universal value, in the excluded volume
limit, of $\eta_p/\eta_{p,0}$, at $\beta=6$. As in the case of
extrapolated values in the crossover regime, insufficiency in the
accumulated finite chain data is revealed when curves for
different values of $z^*$ fail to extrapolate to a common point.
In the present instance, this typically occurs for values of
characteristic shear rate $\beta \gtrapprox 10$. Extrapolated data
obtained in this manner, at various values of $\beta$ (up to
$\beta \approx 10$), for both the viscometric ratios, are
displayed in Figs.~\ref{evpsiextra}. The horizontal lines,
displayed in Figs.~\ref{crossev} earlier, have also been obtained
similarly.

Both the viscometric functions displayed in Figs.~\ref{evpsiextra}
exhibit a similar dependence on $\beta$\textemdash nearly constant
values at small values of $\beta$, followed by a crossover to a
power law dependence at large values of $\beta$. Results obtained
earlier with the Gaussian approximation are also presented in
Figs.~\ref{evpsiextra} for comparison. While the extent of shear
thinning is under-predicted by the Gaussian approximation for the
values of $\beta$ displayed in the figures, it predicts a larger
slope in the power law regime, which implies that the shear
thinning predicted by it will eventually be greater than
\textit{exact} Brownian dynamics simulations.

The Carreau-Yasuda model, which proposes the following expressions
for describing the variation of the viscometric ratios with
$\beta$,
\begin{equation}
\frac{\eta_{\text{p}} }{ \eta_{\text{p}, \, 0}}= \left[ 1+
(a_\eta \, \beta)^{n_\eta} \right]^{\frac{m_\eta}{n_\eta}};
\quad \frac{\Psi_1 }{ \Psi_{1, \, 0}}= \left[ 1+ (a_\psi \,
\beta)^{n_\psi} \right]^{\frac{m_\psi}{n_\psi}}\label{eqcy}
\end{equation}
has been shown previously by Prakash~\cite{pra02} to provide a
good fit to the results of the Gaussian approximation. As
indicated in Figs.~\ref{evpsiextra}, with a suitable choice of the
fitting coefficients $a_\eta$, $m_\eta$, $n_\eta$ etc., it also
leads to a good fit of Brownian dynamics simulation data. The
values of the fitting coefficients used in the two cases are
displayed in Table~\ref{tabcy}. The constant $m$ represents the
slope of the power law region at large values of $\beta$. In line
with the expectation by visual inspection, the value  of $m$ for
the Gaussian approximation is larger than that for \textit{exact}
Brownian dynamics simulations. Using renormalization group
methods, \"Ottinger~\cite{ott89c,ott90} has previously found that
the values, $m_\eta=-0.25$ and $m_\psi=-0.5$, describe the power
law shear rate dependence of ${\eta_p}/{\eta_{p,0}}$, and
$\Psi_1/\Psi_{1,0}$, respectively, in the excluded volume limit.
As is evident from the values of the power law exponents displayed
in Table~\ref{tabcy}, the predictions of Brownian dynamics
simulations are in remarkable agreement with renormalization group
results.

\begin{table}[tb]
  \caption{Values of Carreau-Yasuda model parameters for
  $\eta_p/\eta_{p,0}$ and $\Psi_{1}/\Psi_{1,0}$.
  Comparison of exact Brownian dynamics simulations (BDS),
  with the Gaussian approximation (GA).}\label{tabcy}
  \vskip3pt
  \begin{tabular}{|c||c|c|c|}\hline
    % \multicolumn{4}{|c|}{Brownian Dynamics Simulations}\\ \hline
    & $m_\eta$ &  $n_\eta$ & $a_\eta$  \\ \hline\hline
   BDS & $-0.255\pm 0.009$ & $1.925\pm 0.100$ & $0.712\pm 0.050$ \\ \hline
    GA & $-0.288\pm 0.006 $ & $1.675 \pm 0.097 $ & $0.429\pm 0.025$ \\ \hline\hline
    % \multicolumn{4}{|c|}{Gaussian Approximation}\\ \hline
     &$m_\psi$ &  $n_\psi$    & $a_\psi$ \\ \hline\hline
    BDS & $-0.52\pm 0.02$ & $2.395\pm 0.090$ & $1.312\pm 0.020$  \\ \hline
    GA & $-0.750\pm 0.014$ & $1.878 \pm 0.795  $   & $0.507 \pm 0.012 $ \\ \hline
  \end{tabular}
 \end{table}

The expression for $U_{\Psi\eta}$, Eq.~(\ref{UPE}), can be
rewritten in the form,
\begin{align}
U_{\Psi\eta} & =
U_{\Psi\eta,0} \, \frac{(\Psi_{1}/\Psi_{1,0})}{(\eta_{p}/\eta_{p,0})^2} \label{UPE0}\\
& = U_{\Psi\eta,0} \, \frac{ \left[ 1 + (a_\psi \beta)^{n_\psi}
\right]^{(m_\psi/n_\psi)}}{\left[ 1 + (a_\eta \beta)^{n_\eta}
\right]^{(2m_\eta/n_\eta)}} \label{UPE1}
\end{align}
where, $U_{\Psi\eta,0}$ is the value of $U_{\Psi\eta}$ in the zero
shear rate limit, and the Carreau-Yasuda model expression,
Eq.~(\ref{eqcy}), has been used for the viscometric ratios. Both
renormalization group results, and the present Brownian dynamics
simulations suggest that, $m_\psi \approx 2 m_\eta \approx -0.5$
(see Table~\ref{tabcy}). For $\beta \gg 1$, therefore,
Eq.~(\ref{UPE1}) can be simplified to,
\begin{equation}
U_{\Psi\eta} = U_{\Psi\eta,0} \, \sqrt{\frac{a_\eta}{a_\psi}}
\label{UPE2}
\end{equation}
Kumar and Prakash~\cite{kumpra03} have estimated that in the
excluded volume limit, $U_{\Psi\eta,0}= 0.771 \pm 0.009$.
Substituting this result, and the values for $a_\eta$ and $a_\psi$
from Table~\ref{tabcy} into Eq.~(\ref{UPE2}), and using standard
methods for the estimation of error propagation, leads to
$U_{\Psi\eta}= 0.568 \pm 0.038$. Thus, the universal ratio
$U_{\Psi\eta}$ decreases from a zero shear rate value of $0.771
\pm 0.009$, to a shear rate independent value of $0.568 \pm 0.038$
in the limit of large shear rates.

\begin{figure}[t]
    \resizebox{8.5cm}{!}{\includegraphics*{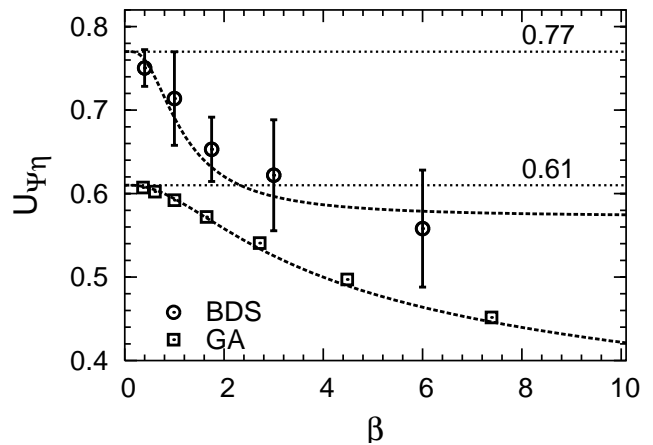} }
    \caption{The dependence of the universal ratio
    $U_{\Psi\eta}$ on the characteristic shear rate $\beta$.
    The open circles are predictions of Brownian dynamics
    simulations, while the squares are predictions of
    the Gaussian approximation, calculated from
    data reported by Prakash.~\cite{pra01b} The dashed curves
    have been obtained from Eq.~(\ref{UPE1}), using values
    for the Carreau-Yasuda model parameters given
    in Table~\ref{tabcy}. The horizontal
    lines represent the zero-shear rate predictions in the
    excluded volume limit, reproduced from
    ref.~\onlinecite{kumpra03}. }
    \label{upefig}
\end{figure}

Figure~\ref{upefig} displays the dependence of $U_{\Psi\eta}$ on
$\beta$, obtained by carrying out Brownian dynamics simulations
(open circles), and with the Gaussian approximation (squares). In
each case, the symbols denote values that have been calculated
using Eq.~(\ref{UPE0}), with the excluded volume limit values
substituted for all the quantities on the right hand side. The
dashed curves have been obtained from Eq.~(\ref{UPE1}), using mean
values for the Carreau-Yasuda model parameters given in
Table~\ref{tabcy}. Unlike in the case of \textit{exact} Brownian
dynamics simulations, since $m_\psi \neq 2 m_\eta$ in the Gaussian
approximation, $U_{\Psi\eta}$ does not reach a constant value in
the limit of large $\beta$.

Perhaps the most celebrated result of the study of excluded volume
interactions in dilute polymer solutions is the observation that
at large molecular weights, the equilibrium root-mean-square
radius of gyration of the polymer coil, $R_\text{g}$, scales with
molecular weight as a power law, $R_\text{g} \sim M^\nu$, where
$\nu$ is a universal exponent, independent of the particular
polymer-solvent system. Pierleoni and Ryckaert~\cite{pierryck00}
have shown recently, by calculating the structure factor for
bead-spring chains with up to 300 beads, that in the presence of
excluded volume interactions, two distinct regimes can be
identified in the structure of a linear polymer chain undergoing
shear flow. At large length scales, the polymer coil has a
Rouse-like behavior, while at small length scales, where the flow
field has not yet distorted the isotropic coil structure, the
polymer coil exhibits typical self-avoiding statistics. In the
light of these observations we expect, therefore, that in the
limit of infinite chain length, the scaling exponent $\nu$ will
remain unaltered from its equilibrium value, since, at any
particular shear rate, more and more polymer length scales will be
smaller than the smallest length scale probed by the flow field.
Recently, Kumar and Prakash~\cite{kumpra03} have obtained the
value of $\nu$ at equilibrium, by exploiting a unique feature of
the solution to the excluded volume problem. We show here that the
same procedure can also be applied to find the value of $\nu$ in
the presence of shear flow. In order to do so, however, it is
necessary to make a few introductory remarks.

While the power law scaling of $R_\text{g}$ with chain length $N$
occurs only in the excluded volume limit, it is common to describe
the dependence of $R_\text{g}$ on $N$, at all values of chain
length, with an apparent power law, $R_\text{g} \sim
N^{\nu_\text{eff}}$, with $\nu_\text{eff}$ representing an
effective exponent that approaches its critical value $\nu$, as $N
\to \infty$. Schafer and coworkers~\cite{grassberger97,schafer}
have shown that the manner in which $\nu_\text{eff}$ approaches
its asymptotic limit is strongly dependent on the magnitude of the
parameter $z^*$ relative to its \textit{fixed point} value
$z^*_\text{f}$. Discussions of the origin of the fixed point, and
its significance can be found in treatises on renormalization
group methods.~\cite{freed,clojan,schafer} Basically, the
existence of power law behavior in the excluded volume limit, such
as the one observed for $R_\text{g}$, is intimately connected to
the existence of a fixed point value for the parameter $z^*$.
Schafer and coworkers have shown, using both renormalization group
methods, and Monte Carlo simulations, that $\nu_\text{eff} \to
\nu$ on two distinct branches, (i) the strong-coupling branch
corresponding to $z^* > z^*_\text{f}$, and (ii) the weak-coupling
branch corresponding to $z^* < z^*_\text{f}$. On the weak-coupling
branch, for increasing values of $N$ (or equivalently $z$),
$\nu_\text{eff}$ approaches $\nu$ from below, increasing rapidly
at first before approaching the asymptotic value very gradually.
On the other hand, on the strong-coupling branch, $\nu_\text{eff}$
approaches $\nu$ from above, decreasing rapidly before approaching
the asymptotic value slowly. Kumar and Prakash~\cite{kumpra03}
have shown that Brownian dynamics simulations are able to capture
the existence of the dual branched structure of the solution to
the excluded volume problem. Furthermore, they show that this
distinctive feature of the solution implies that at fixed values
of $z$ and $d^*$, a plot of $\nu_\text{eff}$ versus $z^*$, for a
range of values spanning the fixed point, has a characteristic
shape exhibiting a marked inflection point, from which one can
obtain both the fixed point $z^*_\text{f}$, and the critical
exponent $\nu$ (see Fig.~11 in ref.~\onlinecite{kumpra03}). In
particular, they found that at equilibrium, $\nu=0.6$, and $0.28
\le z^*_\text{f} \le 0.3$.

\begin{figure}[tb]
  \resizebox{8.5cm}{!}{\includegraphics*{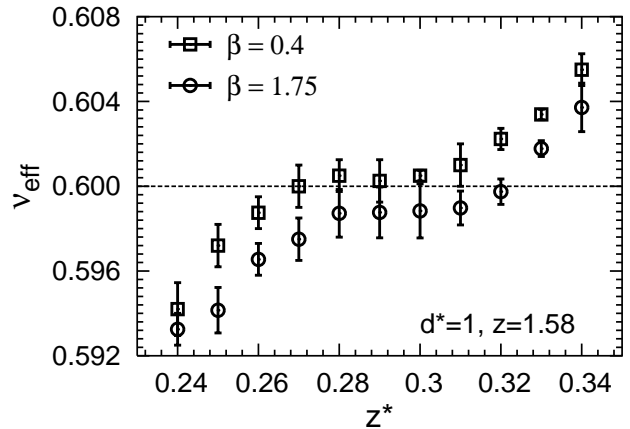} }
  \caption{The effective exponent $\nu_\text{eff}$ as a function of the
strength of excluded volume interactions $z^*$, predicted by
Brownian dynamics simulations, at two different vales of the shear
rate $\beta$.}
  \label{inflxn}
\end{figure}

By collecting data on $R_\text{g}$ as a function of $N$, at
various constant values of the parameters $z^*$, and $\beta$, the
value of $\nu_\text{eff}$, at particular values of $z$ and $d^*$,
can be calculated with the help of the expression, $
\nu_\text{eff}= {\partial\ln R_\text{g}}/{\partial\ln N}$. The
symbols in Fig.~\ref{inflxn} have been obtained in this manner,
for two representative values of the characteristic shear rate
$\beta$. It is clear from the figure that the dual branched
structure of the solution persists into the non-equilibrium
regime, and the characteristic shape of the dependence of
$\nu_\text{eff}$ on $z^*$, observed at equilibrium, is preserved
at finite values of the shear rate. The point of inflection, for
the data corresponding to $\beta = 0.4$, clearly indicates that
both $z^*_\text{f}$ and $\nu$ are unaltered from their equilibrium
values. At $\beta = 1.75$, while $z^*_\text{f}$ lies in the same
interval as at equilibrium, there is a small decrease in the value
of $\nu$, arising perhaps due the relatively small values of $N$
used to acquire the data. The results displayed in
Fig.~\ref{inflxn} are consistent with the argument that, for
chains having sufficiently large number of beads, regardless of
the characteristic shear rate $\beta$, there exists a local length
scale below which the beads do not experience flow, and are
effectively at equilibrium.

It is appropriate now to discuss the earlier observation in
Fig.~\ref{crossev} that the \textit{exact} crossover functions for
${\eta_p}/{\eta_{p,0}}$ and $\Psi_1/\Psi_{1,0}$, in the limit of
large $z$, do not approach the asymptotic predictions obtained in
the excluded volume limit. Kumar and Prakash~\cite{kumpra03} have
speculated, in connection with a similar observation made earlier
for universal equilibrium and linear viscoelastic ratios, that the
origin of this behavior might lie in the fundamental difference
between systems described by the crossover region, and those that
correspond to the excluded volume limit. As pointed out earlier,
since the crossover behavior has been obtained by keeping $z$
constant as $N \to \infty$ (which implies that $z^* \to 0$, and $T
\to T_\theta$), points on a crossover curve correspond to systems,
(i) that are infinitesimally close to the $\theta$-temperature,
and (ii) that satisfy the criteria for belonging to the
weak-coupling branch. On the other hand, since the excluded volume
limit behavior has been obtained for systems with a finite value
of $z^*$, they correspond to systems, (i) with a non-zero
difference between $T$ and $T_\theta$, and (ii) that belong to
either the weak or the strong-coupling branch. Interestingly, it
was shown in ref.~\onlinecite{kumpra03} that, at equilibrium, the
Gaussian approximation does not possess the dual-branched
structure of the exact solution. We have not examined here whether
such is the case even at finite shear rates. Nevertheless, as
pointed out previously, both at equilibrium and at steady state,
the large $z$ crossover behavior of the Gaussian approximation
coincides with the behavior in the excluded volume limit, unlike
in the case of the exact solution.

A number of studies aimed at predicting the orientation angle and
orientation resistance in shear flow, associated with various
tensorial quantities, as a function of the shear rate, have been
carried out
previously.~\cite{bossott95,knuetal93,austetal99,pierryck00,ciftor04}
The general consensus is that these quantities are universal
properties of a dilute polymer solution, in the sense that, at the
same reduced shear rate $\beta$, chains of different lengths have
the same orientation angles and resistances. The universal
dependence of $\chi_G$ and $m_G$ on $\beta$, predicted by the
present Brownian dynamics simulations in the excluded volume
limit, will be presented shortly below. However, the universal
shear rate dependence of the orientation angle $\chi_\tau$, and
the orientation resistance $m_\tau$ (with the expected power law
behavior at large values of $\beta$), can be derived in a
straightforward manner from the results that have been obtained
above for the shear rate dependence of the viscometric functions
in the excluded volume limit, namely, Eqs.~(\ref{eqcy}).
\begin{figure}[htb]
  \resizebox{8.5cm}{!}{\includegraphics*{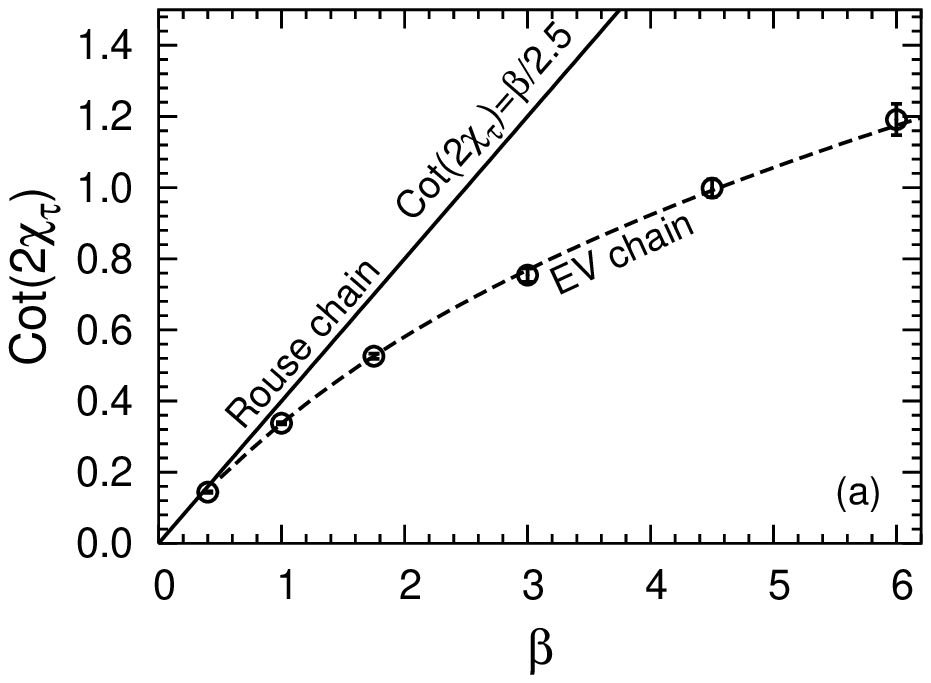} }
  \resizebox{8.5cm}{!}{\includegraphics*{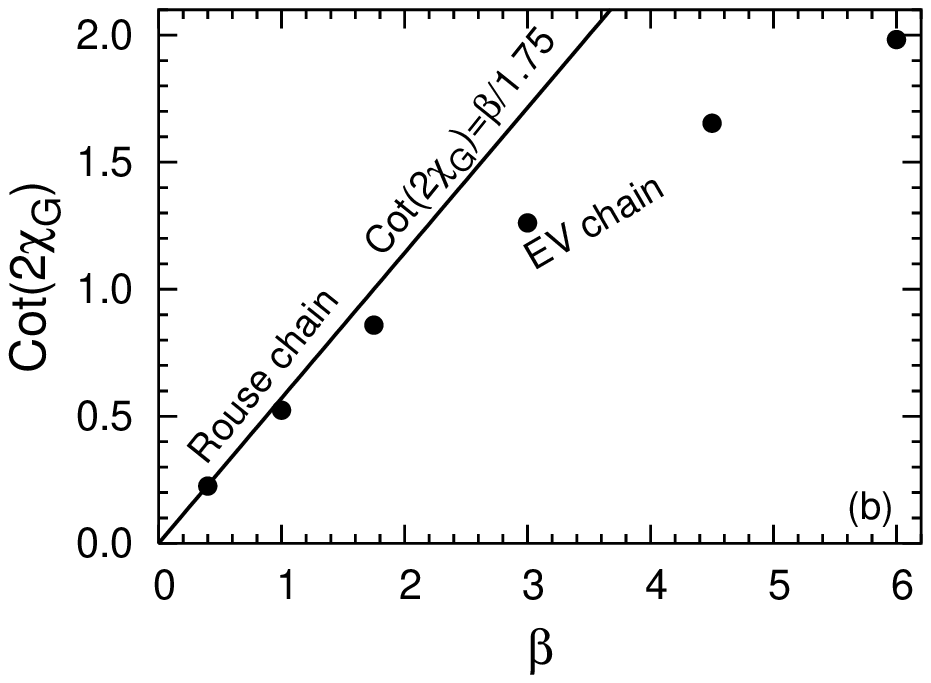} }
  \caption{Universal dependence of the cotangent of the orientation
  angles, (a) $\chi_\tau$ (open circles), and (b) $\chi_G$ (filled circles), on
  the characteristic shear rate $\beta$, in the excluded
  volume limit, obtained by Brownian dynamics simulations.
  The dashed curve through the $\chi_\tau$ data has been drawn using
  Eq.~(\ref{chitau2}), with parameter values reported in
Table~\ref{tabcy}. The solid lines are the predictions of the
Rouse model.} \label{chifigs}
\end{figure}

Substituting the expressions for the viscometric functions,
Eqs.~(\ref{sfvis}), into the defining expressions for $\chi_\tau$
and $m_\tau$, Eq.~(\ref{chitau}), and using the Carreau-Yasuda
model, and the relation between $\dot \gamma$ and $\beta$, leads
to the following expressions,
\begin{align}
\cot(2 \chi_\tau) & = \frac{U_{\Psi\eta,0}}{2}
\frac{(\Psi_{1}/\Psi_{1,0})}{(\eta_{p}/\eta_{p,0})} \,
\beta \label{chitau1}\\
& = \frac{U_{\Psi\eta,0}}{2}\frac{ \left[ 1 + (a_\psi
\beta)^{n_\psi} \right]^{(m_\psi/n_\psi)}}{\left[ 1 +
(a_\eta \beta)^{n_\eta} \right]^{(m_\eta/n_\eta)}} \, \beta
\label{chitau2} \\
%\nonumber\\
m_\tau & =
\frac{2}{U_{\Psi\eta,0}}\frac{(\eta_{p}/\eta_{p,0})}{
(\Psi_{1}/\Psi_{1,0})} \label{mtau1}\\
& = \frac{2}{U_{\Psi\eta,0}}\frac{\left[ 1 + (a_\eta
\beta)^{n_\eta} \right]^{(m_\eta/n_\eta)}}{ \left[ 1 +
(a_\psi \beta)^{n_\psi}
\right]^{(m_\psi/n_\psi)}}\label{mtau2}
\end{align}
On using the relation $m_\psi=2m_\eta$,
Eqs.~(\ref{chitau2}) and~(\ref{mtau2}) simplify to the
following power law dependencies of $\cot(2\chi_\tau)$ and
$m_\tau$ on $\beta$, in the limit of large $\beta$,
\begin{align}
\cot(2 \chi_\tau) & = \frac{U_{\Psi\eta,0}}{2}
\left(\frac{a_\psi^2}{a_\eta}\right)^{m_\eta} \beta^{1+m_\eta}\label{chitau3}\\
m_\tau & =
\frac{2}{U_{\Psi\eta,0}}\left(\frac{a_\eta}{a_\psi^2}\right)^{m_\eta}
\beta^{-m_\eta}\label{mtau3}
\end{align}
Since both renormalization group theory,~\cite{ott89c,ott90} and
the current Brownian dynamics simulations suggest $m_\eta \approx
-0.25$ (see Table~\ref{tabcy}), Eqs.~(\ref{chitau3})
and~(\ref{mtau3}) indicate that $\cot(2 \chi_\tau)$ grows
approximately as $\beta^{0.75}$, while $m_\tau$ has a weaker
dependence, growing approximately as $\beta^{0.25}$, in the limit
of large $\beta$. These predictions are of course restricted to
the influence of excluded volume interactions on these quantities,
and a thorough comparison with experiment requires the
incorporation of hydrodynamic interaction effects. Bossart and
\"Ottinger~\cite{bossott95} and Cifre and de la
Torre~\cite{ciftor04} have shown previously that the presence of
hydrodynamic interactions also leads to shear rate dependent
orientation angles and resistances. Interestingly, however, in
contrast to the prediction obtained here in the presence of
excluded volume interactions alone, they find that at large values
of $\beta$, the orientation resistance $m_\tau$ \textit{decreases}
with increasing shear rate. As is well known, the presence of
hydrodynamic interactions leads to the prediction of \textit{shear
thickening} at large values of $\beta$.~\cite{pra99} It is clear
from Eq.~(\ref{mtau3}), that a positive value of the exponent
$m_\eta$, would explain the observation of Bossart and \"Ottinger,
and Cifre and de la Torre.

Figures~\ref{chifigs} display the shear rate dependence of the
orientation angles obtained here by Brownian dynamics simulations
in the excluded volume limit. In the case of $\cot(2 \chi_\tau)$,
values for ${\eta_p}/{\eta_{p,0}}$ and $\Psi_1/\Psi_{1,0}$,
obtained at various values of $\beta$, in the excluded volume
limit, have been substituted on the right hand side of
Eqs.~(\ref{chitau1}), with $U_{\Psi\eta,0}=0.771 \pm 0.009$. On
the other hand, $\cot(2 \chi_G)$ has been evaluated by
accumulating finite chain data using Eq.~(\ref{chiG}), at a
constant value of $z^* =0.28$ (the fixed point value), followed by
extrapolation to the long chain limit. Comparison with the Rouse
model (which predicts a linear dependence on $\beta$ in each
case), clearly indicates that the presence of excluded volume
interactions leads to a weakening of the alignment of the polymer
coil in the flow direction. This has been observed previously both
theoretically,~\cite{pierryck00} and
experimentally.~\cite{leemul99} Since, at the same shear rate
$\beta$, $\cot(2 \chi_G) > \cot(2 \chi_\tau)$, the tensor $\bfG$
is more easily oriented than $\btau^\text{p}$.

\begin{figure}[tb]
  \resizebox{8.5cm}{!}{\includegraphics*{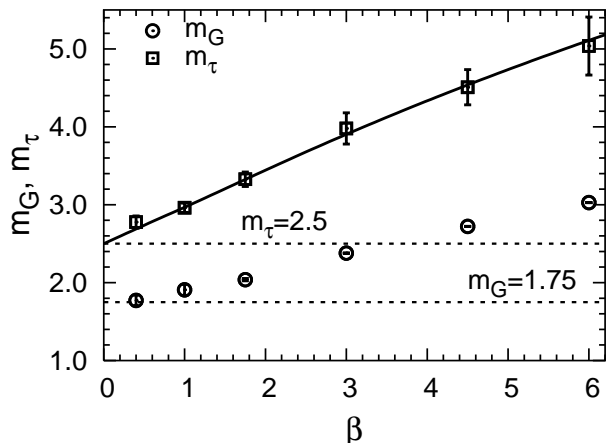} }
  \caption{Universal dependence of the orientation resistances
  $m_\tau$ (squares) and $m_G$ (circles), on
  the characteristic shear rate $\beta$, in the excluded
  volume limit. The line through the $m_\tau$ data has been drawn using
  Eq.~(\ref{mtau2}), with parameter values reported in
Table~\ref{tabcy}. The horizontal lines are the predictions of the
Rouse model.} \label{mfigs}
\end{figure}

Figure~\ref{mfigs} displays the universal dependence of the
orientation resistances $m_G$ and $m_\tau$, in the excluded volume
limit, on $\beta$. As in the case of $\chi_\tau$, the orientation
resistance $m_\tau$ has been obtained by substituting known values
on the right hand side of Eq.~(\ref{mtau1}). On the other hand,
$m_G$ has been obtained from the Brownian dynamics simulation data
for $\cot(2 \chi_G)$, displayed in Fig.~\ref{chifigs}(b), by using
the defining expression, $m_G = \beta / \cot(2 \chi_G)$. It is
evident from the figure that both $m_G$ and $m_\tau$ increase with
$\beta$, indicating that it gets increasingly difficult to orient
the polymer coil with increasing shear rate. This has been
previously predicted,~\cite{pierryck95} and observed
experimentally at relatively low values of
$\beta$.~\cite{linspr93} The fact that $\bfG$ is more easily
oriented than $\btau^\text{p}$ can also be seen from the behavior
of the orientation resistances $m_G$ and $m_\tau$, since $m_\tau >
m_G$, for all values of $\beta$. As pointed out previously by
Bossart and \"Ottinger,~\cite{bossott95} flow birefringence
experiments (which measure $\chi_\tau$), reflect orientational
ordering on length scales smaller than those probed by light
scattering experiments (which measure $\chi_G$). Since smaller
length scales are less affected by flow, they exhibit greater
resistance to orientational ordering, leading to the observed
relative magnitudes of $m_G$ and $m_\tau$.

It is of particular interest to examine the universal
values of the orientation resistances in the limit of zero
shear rate, namely, $m_{G,0}$ and $m_{\tau,0}$, since a
number of studies have been confined to the low shear rate
regime. Bossart and \"Ottinger,~\cite{bossott95} and
Pierleoni and Ryckaert~\cite{pierryck95} have argued
previously that the effect of excluded volume interactions
on the orientation resistance is negligible. Bossart and
\"Ottinger,~\cite{bossott97} have found, from careful
experiments, that in a good solvent (polystyrene in
bromobenzene), $m_{\tau,0}=3.74 \pm 0.28$, compared to the
value of $3.13 \pm 0.42$ in a theta solvent (polystyrene in
4-bromo-$\alpha$-benzyl alcohol). Given the magnitude of
the error bars, it can only be concluded that $m_{\tau,0}$
is probably \textit{larger} in a good solvent than in a
theta solvent. It must be borne in mind that both these
values reflect the influence of hydrodynamic interactions.
In the present instance, where only the influence of
excluded volume interactions is considered, the value of
$m_{\tau,0}$ can be obtained from the following expression,
\begin{equation}
m_{\tau,0}  =\frac{2}{U_{\Psi\eta,0}}
\end{equation}
which is derivable from Eq.~(\ref{mtau1}) in the limit of zero
shear rate. Using the previously established \textit{exact}
Brownian dynamics simulations value of $U_{\Psi\eta,0}=0.771 \pm
0.009$,~\cite{kumpra03} leads to, $m_{\tau,0}=2.594 \pm 0.030$,
which is marginally larger than the Rouse value of 2.5. Using
non-equilibrium molecular dynamics simulations of bead-spring
chains with FENE springs, \citet{austetal99} have estimated
$m_{\tau,0}=2.4 \pm 0.3$. Interestingly, in spite of the fact that
the solvent is treated explicitly in these simulations, and
consequently hydrodynamic interactions are taken into account, the
predicted value is not significantly different from the pure
excluded volume prediction obtained here. It is worth noting that
Bossart and \"Ottinger,~\cite{bossott95} who used the Gaussian
approximation, and de la Torre and
co-workers,\cite{knuetal93,ciftor04} who used Brownian dynamics
simulations with hydrodynamic interactions, report the values:
$m_{\tau,0}\approx 3.57$, and $m_{\tau,0}\approx 3.5$,
respectively\textemdash both of which are substantially different
from the prediction of the Rouse model. While an improvement in
solvent quality seems to increase $m_{\tau,0}$ fractionally, our
results seem to suggest that it \textit{decreases} $m_{G,0}$
fractionally. By fitting the $m_G$ data in Fig.~\ref{mfigs} with a
least squares third order polynomial, and extrapolating to $\beta
\to 0$, we find that in the excluded volume limit, $m_{G,0}=1.69
\pm 0.03$, which is slightly smaller than the Rouse prediction of
$m_{G,0}=1.75$. The value obtained here is in good agreement with
the explicit solvent non-equilibrium molecular dynamics
simulations of \citet{austetal99} and the results of
renormalization group calculations,~\cite{bossott95} both of which
predict $m_{G,0}=1.7$. Since these studies account for
hydrodynamic interactions, it is clear that hydrodynamic
interactions have no effect on $m_{G,0}$. Pierleoni and Ryckaert
have also observed a similar pattern when comparing their explicit
solvent molecular dynamics simulations~\cite{pierryck95} with
their Brownian dynamics simulations (in which only excluded volume
interactions were taken into account).~\cite{pierryck00} It is
well known that hydrodynamic interactions have a significant
influence on \textit{dynamic} properties, but have no influence on
\textit{static} properties, since the equilibrium distribution
function is unaltered in their presence. This suggests, perhaps
not surprisingly, that $m_{G,0}$ is static property, while
$m_{\tau,0}$ is a dynamic property.

\section{Conclusions \label{end}}

Results of a detailed Brownian dynamics simulations study,
of a polymer solution undergoing steady shear flow, have
been presented. The polymer molecule has been modelled by a
bead-spring chain, and a narrow Gaussian excluded volume
potential, that acts between pairs of beads, has been used
to mimic the influence of solvent quality.

Material properties have been shown to become independent of the
range of excluded volume interactions $d^*$, and the number of
beads $N$, in the limit of large $N$. Furthermore, it has been
found that master plots are obtained when, (i) the influence of
the strength of excluded volume interactions is interpreted in
terms of the solvent quality parameter $z = z^* \sqrt{N}$, and
(ii) the dependence on the shear rate $\dot \gamma$ is interpreted
in terms of the characteristic non-dimensional shear rate,
$\beta$. In this work, we have explored the universal dependence
of the non-dimensional viscosity ratio $(\eta_{\text{p}} /
\eta_{\text{p}, \, 0})$, the non-dimensional first normal stress
difference ratio $(\Psi_1 / \Psi_{1, \, 0})$, the orientation
angles $\chi_\tau$ and $\chi_G$, and the orientation resistances
$m_\tau$ and $m_G$, on the characteristic shear rate $\beta$ and
the solvent quality $z$.

The extent of shear thinning has been found to increase as $z$
increases (Figs.~\ref{psiextra}). However, the incremental
increase in shear thinning saturates as $z$ increases
(Figs.~\ref{crossev}). The existence of universal viscometric
ratio versus shear rate curves, independent of all model
parameters, in the excluded volume limit, $z \to \infty$, has been
verified (Figs.~\ref{evpsiextra}). The shear rate dependence of
both the universal viscometric ratios has been found to be well
described by the Carreau-Yasuda model, with power law decay
exponents at large values of $\beta$, that are in close agreement
with values predicted by renormalization group theory
(Table.~\ref{tabcy}).

The asymptotic values of the exact crossover functions for
$(\eta_{\text{p}} / \eta_{\text{p}, \, 0})$ and $(\Psi_1 /
\Psi_{1, \, 0})$, in the limit of large $z$, are found not to
approach the universal predictions obtained in the excluded volume
limit, unlike previous predictions of the Gaussian approximation
(Figs.~\ref{crossev}).

The shear rate dependence of the universal ratio $U_{\Psi\eta}$
has been obtained in the excluded volume limit. The ratio is found
to decrease from the zero shear rate value of $0.771 \pm 0.009$,
to a constant value of $0.568 \pm 0.038$, in the limit of large
shear rates (Fig.~\ref{upefig}).

The dual branched structure of the solution, elucidated recently
by Sc\"afer and co-workers,~\cite{grassberger97,schafer} and shown
to be captured by Brownian dynamics simulations at
equilibrium,~\cite{kumpra03} has been found to persist into the
non-equilibrium regime. As in the equilibrium case, this
distinctive structure has been exploited here to obtain an
estimate of both the fixed point of the strength of excluded
volume interactions, and the critical exponent $\nu$. Both
quantities are found to remain unaltered in the presence of shear
flow (Fig.~\ref{inflxn}).

The universal dependence of $\chi_\tau$, $\chi_G$, $m_\tau$ and
$m_G$, on $\beta$, in the excluded volume limit, has been
obtained. The presence of excluded volume interactions has been
shown to weaken the alignment of polymer coils in the flow
direction (Figs.~\ref{chifigs}). Of the two tensors, the radius of
gyration tensor $\bfG$ is found to be more easily oriented than
$\btau^\text{p}$ (Fig.~\ref{mfigs}). The power law shear rate
dependence of $\chi_\tau$ ($\sim \beta^{0.75}$) and $m_\tau$
($\sim \beta^{0.25}$), in the limit of large shear rate, has been
obtained by exploiting the Carreau-Yasuda model fit of the
universal viscometric functions. The presence of excluded volume
interactions leads to a marginal increase in $m_{\tau,0}$, while
marginally reducing $m_{G,0}$.
%\vskip40pt
\begin{acknowledgments}
We gratefully acknowledge the Victorian Partnership for
Advanced Computing (VPAC) for a grant under the Expertise
program, and both VPAC and the Australian Partnership for
Advanced Computing (APAC) for the use of their
computational facilities.
\end{acknowledgments}

\bibliography{513431JCP}

\begin{thebibliography}{41}
\expandafter\ifx\csname natexlab\endcsname\relax\def\natexlab#1{#1}\fi
\expandafter\ifx\csname bibnamefont\endcsname\relax
  \def\bibnamefont#1{#1}\fi
\expandafter\ifx\csname bibfnamefont\endcsname\relax
  \def\bibfnamefont#1{#1}\fi
\expandafter\ifx\csname citenamefont\endcsname\relax
  \def\citenamefont#1{#1}\fi
\expandafter\ifx\csname url\endcsname\relax
  \def\url#1{\texttt{#1}}\fi
\expandafter\ifx\csname urlprefix\endcsname\relax\def\urlprefix{URL }\fi
\providecommand{\bibinfo}[2]{#2}
\providecommand{\eprint}[2][]{\url{#2}}

\bibitem[{\citenamefont{Miyaki and Fujita}(1981)}]{miyfuj81}
\bibinfo{author}{\bibfnamefont{Y.}~\bibnamefont{Miyaki}} \bibnamefont{and}
  \bibinfo{author}{\bibfnamefont{H.}~\bibnamefont{Fujita}},
  \bibinfo{journal}{Macromolecules} \textbf{\bibinfo{volume}{14}},
  \bibinfo{pages}{742} (\bibinfo{year}{1981}).

\bibitem[{\citenamefont{Vidakovic and Rondelez}(1985)}]{vidron85}
\bibinfo{author}{\bibfnamefont{P.}~\bibnamefont{Vidakovic}} \bibnamefont{and}
  \bibinfo{author}{\bibfnamefont{F.}~\bibnamefont{Rondelez}},
  \bibinfo{journal}{Macromolecules} \textbf{\bibinfo{volume}{18}},
  \bibinfo{pages}{700} (\bibinfo{year}{1985}).

\bibitem[{\citenamefont{Hayward and Graessley}(1999)}]{haygra99}
\bibinfo{author}{\bibfnamefont{R.~C.} \bibnamefont{Hayward}} \bibnamefont{and}
  \bibinfo{author}{\bibfnamefont{W.~W.} \bibnamefont{Graessley}},
  \bibinfo{journal}{Macromolecules} \textbf{\bibinfo{volume}{32}},
  \bibinfo{pages}{3502} (\bibinfo{year}{1999}).

\bibitem[{\citenamefont{Bercea et~al.}(1999)\citenamefont{Bercea, Ioan, Ioan,
  Simionescu, and Simionescu}}]{beretal99}
\bibinfo{author}{\bibfnamefont{M.}~\bibnamefont{Bercea}},
  \bibinfo{author}{\bibfnamefont{C.}~\bibnamefont{Ioan}},
  \bibinfo{author}{\bibfnamefont{S.}~\bibnamefont{Ioan}},
  \bibinfo{author}{\bibfnamefont{B.~C.} \bibnamefont{Simionescu}},
  \bibnamefont{and} \bibinfo{author}{\bibfnamefont{C.~I.}
  \bibnamefont{Simionescu}}, \bibinfo{journal}{Progress in Polymer Science}
  \textbf{\bibinfo{volume}{24}}, \bibinfo{pages}{379} (\bibinfo{year}{1999}).

\bibitem[{\citenamefont{Doi and Edwards}(1986)}]{doiedw}
\bibinfo{author}{\bibfnamefont{M.}~\bibnamefont{Doi}} \bibnamefont{and}
  \bibinfo{author}{\bibfnamefont{S.~F.} \bibnamefont{Edwards}},
  \emph{\bibinfo{title}{The Theory of Polymer Dynamics}}
  (\bibinfo{publisher}{Oxford University Press}, \bibinfo{address}{New York},
  \bibinfo{year}{1986}).

\bibitem[{\citenamefont{Freed}(1987)}]{freed}
\bibinfo{author}{\bibfnamefont{K.~F.} \bibnamefont{Freed}},
  \emph{\bibinfo{title}{Renormalization Group Theory of Macromolecules}}
  (\bibinfo{publisher}{Wiley}, \bibinfo{address}{New York},
  \bibinfo{year}{1987}).

\bibitem[{\citenamefont{des Cloizeaux and Jannink}(1990)}]{clojan}
\bibinfo{author}{\bibfnamefont{J.}~\bibnamefont{des Cloizeaux}}
  \bibnamefont{and} \bibinfo{author}{\bibfnamefont{G.}~\bibnamefont{Jannink}},
  \emph{\bibinfo{title}{Polymers in Solution, Their Modeling and Structure}}
  (\bibinfo{publisher}{Oxford Science Publishers}, \bibinfo{year}{1990}).

\bibitem[{\citenamefont{Sch{\"{a}}fer}(1999)}]{schafer}
\bibinfo{author}{\bibfnamefont{L.}~\bibnamefont{Sch{\"{a}}fer}},
  \emph{\bibinfo{title}{Excluded Volume Effects in Polymer Solutions}}
  (\bibinfo{publisher}{Springer-Verlag}, \bibinfo{address}{Berlin},
  \bibinfo{year}{1999}).

\bibitem[{\citenamefont{{\"{O}}ttinger}(1989)}]{ott89c}
\bibinfo{author}{\bibfnamefont{H.~C.} \bibnamefont{{\"{O}}ttinger}},
  \bibinfo{journal}{Phys. Rev. A} \textbf{\bibinfo{volume}{40}},
  \bibinfo{pages}{2664} (\bibinfo{year}{1989}).

\bibitem[{\citenamefont{Zylka and {\"{O}}ttinger}(1991)}]{zylott91}
\bibinfo{author}{\bibfnamefont{W.}~\bibnamefont{Zylka}} \bibnamefont{and}
  \bibinfo{author}{\bibfnamefont{H.~C.} \bibnamefont{{\"{O}}ttinger}},
  \bibinfo{journal}{Macromolecules} \textbf{\bibinfo{volume}{24}},
  \bibinfo{pages}{484} (\bibinfo{year}{1991}).

\bibitem[{\citenamefont{Andrews et~al.}(1998)\citenamefont{Andrews, Doufas, and
  McHugh}}]{andetal98}
\bibinfo{author}{\bibfnamefont{N.~C.} \bibnamefont{Andrews}},
  \bibinfo{author}{\bibfnamefont{A.~K.} \bibnamefont{Doufas}},
  \bibnamefont{and} \bibinfo{author}{\bibfnamefont{A.~J.}
  \bibnamefont{McHugh}}, \bibinfo{journal}{Macromolecules}
  \textbf{\bibinfo{volume}{31}}, \bibinfo{pages}{3104} (\bibinfo{year}{1998}).

\bibitem[{\citenamefont{Cifre and de~la Torre}(1999)}]{ciftor99}
\bibinfo{author}{\bibfnamefont{J.~G.~H.} \bibnamefont{Cifre}} \bibnamefont{and}
  \bibinfo{author}{\bibfnamefont{J.~G.} \bibnamefont{de~la Torre}},
  \bibinfo{journal}{J. Rheol.} \textbf{\bibinfo{volume}{43}},
  \bibinfo{pages}{339} (\bibinfo{year}{1999}).

\bibitem[{\citenamefont{Prakash and {\"{O}}ttinger}(1999)}]{praott99}
\bibinfo{author}{\bibfnamefont{J.~R.} \bibnamefont{Prakash}} \bibnamefont{and}
  \bibinfo{author}{\bibfnamefont{H.~C.} \bibnamefont{{\"{O}}ttinger}},
  \bibinfo{journal}{Macromolecules} \textbf{\bibinfo{volume}{32}},
  \bibinfo{pages}{2028} (\bibinfo{year}{1999}).

\bibitem[{\citenamefont{Li and Larson}(2000)}]{lilar00a}
\bibinfo{author}{\bibfnamefont{L.}~\bibnamefont{Li}} \bibnamefont{and}
  \bibinfo{author}{\bibfnamefont{R.~G.} \bibnamefont{Larson}},
  \bibinfo{journal}{Rheol. Acta} \textbf{\bibinfo{volume}{39}},
  \bibinfo{pages}{419} (\bibinfo{year}{2000}).

\bibitem[{\citenamefont{Pierleoni and Ryckaert}(2000)}]{pierryck00}
\bibinfo{author}{\bibfnamefont{C.}~\bibnamefont{Pierleoni}} \bibnamefont{and}
  \bibinfo{author}{\bibfnamefont{J.-P.} \bibnamefont{Ryckaert}},
  \bibinfo{journal}{J. Chem. Phys.} \textbf{\bibinfo{volume}{113}},
  \bibinfo{pages}{5545} (\bibinfo{year}{2000}).

\bibitem[{\citenamefont{Jendrejack et~al.}(2002)\citenamefont{Jendrejack,
  Graham, and de~Pablo}}]{jenetal02}
\bibinfo{author}{\bibfnamefont{R.~M.} \bibnamefont{Jendrejack}},
  \bibinfo{author}{\bibfnamefont{M.~D.} \bibnamefont{Graham}},
  \bibnamefont{and} \bibinfo{author}{\bibfnamefont{J.~J.}
  \bibnamefont{de~Pablo}}, \bibinfo{journal}{J. Chem. Phys.}
  \textbf{\bibinfo{volume}{116}}, \bibinfo{pages}{7752} (\bibinfo{year}{2002}).

\bibitem[{\citenamefont{Prakash}(2001{\natexlab{a}})}]{pra01a}
\bibinfo{author}{\bibfnamefont{J.~R.} \bibnamefont{Prakash}},
  \bibinfo{journal}{Chem. Eng. Sci.} \textbf{\bibinfo{volume}{56}},
  \bibinfo{pages}{5555} (\bibinfo{year}{2001}{\natexlab{a}}).

\bibitem[{\citenamefont{Prakash}(2001{\natexlab{b}})}]{pra01b}
\bibinfo{author}{\bibfnamefont{J.~R.} \bibnamefont{Prakash}},
  \bibinfo{journal}{Macromolecules} \textbf{\bibinfo{volume}{34}},
  \bibinfo{pages}{3396} (\bibinfo{year}{2001}{\natexlab{b}}).

\bibitem[{\citenamefont{Prakash}(2002)}]{pra02}
\bibinfo{author}{\bibfnamefont{J.~R.} \bibnamefont{Prakash}},
  \bibinfo{journal}{J. Rheol.} \textbf{\bibinfo{volume}{46}},
  \bibinfo{pages}{1353} (\bibinfo{year}{2002}).

\bibitem[{\citenamefont{Prabhakar and Prakash}(2002)}]{prapra02}
\bibinfo{author}{\bibfnamefont{R.}~\bibnamefont{Prabhakar}} \bibnamefont{and}
  \bibinfo{author}{\bibfnamefont{J.~R.} \bibnamefont{Prakash}},
  \bibinfo{journal}{J. Rheol.} \textbf{\bibinfo{volume}{46}},
  \bibinfo{pages}{1191} (\bibinfo{year}{2002}).

\bibitem[{\citenamefont{Prabhakar and Prakash}(2004)}]{prapra04}
\bibinfo{author}{\bibfnamefont{R.}~\bibnamefont{Prabhakar}} \bibnamefont{and}
  \bibinfo{author}{\bibfnamefont{J.~R.} \bibnamefont{Prakash}},
  \bibinfo{journal}{J. Non-Newtonian Fluid Mech.}
  \textbf{\bibinfo{volume}{116}}, \bibinfo{pages}{163} (\bibinfo{year}{2004}).

\bibitem[{\citenamefont{Cifre and de~la Torre}(2004)}]{ciftor04}
\bibinfo{author}{\bibfnamefont{J.~G.~H.} \bibnamefont{Cifre}} \bibnamefont{and}
  \bibinfo{author}{\bibfnamefont{J.~G.} \bibnamefont{de~la Torre}},
  \bibinfo{journal}{Macromol. Theory Simul.} \textbf{\bibinfo{volume}{13}},
  \bibinfo{pages}{273} (\bibinfo{year}{2004}).

\bibitem[{\citenamefont{{\"{O}}ttinger}(1990)}]{ott90}
\bibinfo{author}{\bibfnamefont{H.~C.} \bibnamefont{{\"{O}}ttinger}},
  \bibinfo{journal}{Phys. Rev. A} \textbf{\bibinfo{volume}{41}},
  \bibinfo{pages}{4413} (\bibinfo{year}{1990}).

\bibitem[{\citenamefont{Kumar and Prakash}(2003)}]{kumpra03}
\bibinfo{author}{\bibfnamefont{K.~S.} \bibnamefont{Kumar}} \bibnamefont{and}
  \bibinfo{author}{\bibfnamefont{J.~R.} \bibnamefont{Prakash}},
  \bibinfo{journal}{Macromolecules} \textbf{\bibinfo{volume}{36}},
  \bibinfo{pages}{7842} (\bibinfo{year}{2003}).

\bibitem[{\citenamefont{Cascales and de~la Torre}(1991)}]{castor91}
\bibinfo{author}{\bibfnamefont{J.~J.~L.} \bibnamefont{Cascales}}
  \bibnamefont{and} \bibinfo{author}{\bibfnamefont{J.~G.} \bibnamefont{de~la
  Torre}}, \bibinfo{journal}{Polymer} \textbf{\bibinfo{volume}{32}},
  \bibinfo{pages}{3359} (\bibinfo{year}{1991}).

\bibitem[{\citenamefont{Knudsen et~al.}(1996)\citenamefont{Knudsen, de~la
  Torre, and Elgsaeter}}]{knuetal96}
\bibinfo{author}{\bibfnamefont{K.~D.} \bibnamefont{Knudsen}},
  \bibinfo{author}{\bibfnamefont{J.~G.} \bibnamefont{de~la Torre}},
  \bibnamefont{and}
  \bibinfo{author}{\bibfnamefont{A.}~\bibnamefont{Elgsaeter}},
  \bibinfo{journal}{Polymer} \textbf{\bibinfo{volume}{37}},
  \bibinfo{pages}{1317} (\bibinfo{year}{1996}).

\bibitem[{\citenamefont{Jendrejack et~al.}(2000)\citenamefont{Jendrejack,
  Graham, and de~Pablo}}]{jenetal00}
\bibinfo{author}{\bibfnamefont{R.~M.} \bibnamefont{Jendrejack}},
  \bibinfo{author}{\bibfnamefont{M.~D.} \bibnamefont{Graham}},
  \bibnamefont{and} \bibinfo{author}{\bibfnamefont{J.~J.}
  \bibnamefont{de~Pablo}}, \bibinfo{journal}{J. Chem. Phys.}
  \textbf{\bibinfo{volume}{113}}, \bibinfo{pages}{2894} (\bibinfo{year}{2000}).

\bibitem[{\citenamefont{Kr{\"{o}}ger et~al.}(2000)\citenamefont{Kr{\"{o}}ger,
  Alba-P{\'{e}}rez, Laso, and {\"{O}}ttinger}}]{kroetal00}
\bibinfo{author}{\bibfnamefont{M.}~\bibnamefont{Kr{\"{o}}ger}},
  \bibinfo{author}{\bibfnamefont{A.}~\bibnamefont{Alba-P{\'{e}}rez}},
  \bibinfo{author}{\bibfnamefont{M.}~\bibnamefont{Laso}}, \bibnamefont{and}
  \bibinfo{author}{\bibfnamefont{H.~C.} \bibnamefont{{\"{O}}ttinger}},
  \bibinfo{journal}{J. Chem. Phys.} \textbf{\bibinfo{volume}{113}},
  \bibinfo{pages}{4767} (\bibinfo{year}{2000}).

\bibitem[{\citenamefont{{\"{O}}ttinger}(1996)}]{ottbk}
\bibinfo{author}{\bibfnamefont{H.~C.} \bibnamefont{{\"{O}}ttinger}},
  \emph{\bibinfo{title}{Stochastic Processes in Polymeric Fluids}}
  (\bibinfo{publisher}{Springer-Verlag}, \bibinfo{year}{1996}).

\bibitem[{\citenamefont{Bird et~al.}(1987{\natexlab{a}})\citenamefont{Bird,
  Curtiss, Armstrong, and Hassager}}]{birdetal2}
\bibinfo{author}{\bibfnamefont{R.~B.} \bibnamefont{Bird}},
  \bibinfo{author}{\bibfnamefont{C.~F.} \bibnamefont{Curtiss}},
  \bibinfo{author}{\bibfnamefont{R.~C.} \bibnamefont{Armstrong}},
  \bibnamefont{and} \bibinfo{author}{\bibfnamefont{O.}~\bibnamefont{Hassager}},
  \emph{\bibinfo{title}{Dynamics of Polymeric Liquids - Volume 2: Kinetic
  Theory}} (\bibinfo{publisher}{John Wiley}, \bibinfo{address}{New York},
  \bibinfo{year}{1987}{\natexlab{a}}), \bibinfo{edition}{2nd} ed.

\bibitem[{\citenamefont{Bird et~al.}(1987{\natexlab{b}})\citenamefont{Bird,
  Armstrong, and Hassager}}]{birdetal1}
\bibinfo{author}{\bibfnamefont{R.~B.} \bibnamefont{Bird}},
  \bibinfo{author}{\bibfnamefont{R.~C.} \bibnamefont{Armstrong}},
  \bibnamefont{and} \bibinfo{author}{\bibfnamefont{O.}~\bibnamefont{Hassager}},
  \emph{\bibinfo{title}{Dynamics of Polymeric Liquids - Volume 1: Fluid
  Mechanics}} (\bibinfo{publisher}{John Wiley}, \bibinfo{address}{New York},
  \bibinfo{year}{1987}{\natexlab{b}}), \bibinfo{edition}{2nd} ed.

\bibitem[{\citenamefont{Bossart and {\"{O}}ttinger}(1995)}]{bossott95}
\bibinfo{author}{\bibfnamefont{J.}~\bibnamefont{Bossart}} \bibnamefont{and}
  \bibinfo{author}{\bibfnamefont{H.~C.} \bibnamefont{{\"{O}}ttinger}},
  \bibinfo{journal}{Macromolecules} \textbf{\bibinfo{volume}{28}},
  \bibinfo{pages}{5852} (\bibinfo{year}{1995}).

\bibitem[{\citenamefont{Iniesta and de~la Torre}(1990)}]{iniesta90}
\bibinfo{author}{\bibfnamefont{A.}~\bibnamefont{Iniesta}} \bibnamefont{and}
  \bibinfo{author}{\bibfnamefont{J.~G.} \bibnamefont{de~la Torre}},
  \bibinfo{journal}{J. Chem. Phys.} \textbf{\bibinfo{volume}{92}},
  \bibinfo{pages}{2015} (\bibinfo{year}{1990}).

\bibitem[{\citenamefont{Grassberger et~al.}(1997)\citenamefont{Grassberger,
  Sutter, and Sch\"afer}}]{grassberger97}
\bibinfo{author}{\bibfnamefont{P.}~\bibnamefont{Grassberger}},
  \bibinfo{author}{\bibfnamefont{P.}~\bibnamefont{Sutter}}, \bibnamefont{and}
  \bibinfo{author}{\bibfnamefont{L.}~\bibnamefont{Sch\"afer}},
  \bibinfo{journal}{J. Phys. A: Math. Gen.} \textbf{\bibinfo{volume}{30}},
  \bibinfo{pages}{7039} (\bibinfo{year}{1997}).

\bibitem[{\citenamefont{Knudsen et~al.}(1993)\citenamefont{Knudsen, Elgsaeter,
  Cascales, and de~la Torre}}]{knuetal93}
\bibinfo{author}{\bibfnamefont{K.~D.} \bibnamefont{Knudsen}},
  \bibinfo{author}{\bibfnamefont{A.}~\bibnamefont{Elgsaeter}},
  \bibinfo{author}{\bibfnamefont{J.~J.~L.} \bibnamefont{Cascales}},
  \bibnamefont{and} \bibinfo{author}{\bibfnamefont{J.~G.} \bibnamefont{de~la
  Torre}}, \bibinfo{journal}{Macromolecules} \textbf{\bibinfo{volume}{26}},
  \bibinfo{pages}{3851} (\bibinfo{year}{1993}).

\bibitem[{\citenamefont{Aust et~al.}(1999)\citenamefont{Aust, Kr\"oger, and
  Hess}}]{austetal99}
\bibinfo{author}{\bibfnamefont{C.}~\bibnamefont{Aust}},
  \bibinfo{author}{\bibfnamefont{M.}~\bibnamefont{Kr\"oger}}, \bibnamefont{and}
  \bibinfo{author}{\bibfnamefont{S.}~\bibnamefont{Hess}},
  \bibinfo{journal}{Macromolecules} \textbf{\bibinfo{volume}{32}},
  \bibinfo{pages}{5660} (\bibinfo{year}{1999}).

\bibitem[{\citenamefont{Prakash}(1999)}]{pra99}
\bibinfo{author}{\bibfnamefont{J.~R.} \bibnamefont{Prakash}}, in
  \emph{\bibinfo{booktitle}{Advances in flow and rheology of non-{N}ewtonian
  fluids}}, edited by \bibinfo{editor}{\bibfnamefont{D.~A.}
  \bibnamefont{Siginer}}, \bibinfo{editor}{\bibfnamefont{D.~D.}
  \bibnamefont{Kee}}, \bibnamefont{and} \bibinfo{editor}{\bibfnamefont{R.~P.}
  \bibnamefont{Chhabra}} (\bibinfo{publisher}{Elsevier Science},
  \bibinfo{address}{Rheology Series, Amsterdam}, \bibinfo{year}{1999}), pp.
  \bibinfo{pages}{467--517}.

\bibitem[{\citenamefont{Lee and Muller}(1999)}]{leemul99}
\bibinfo{author}{\bibfnamefont{E.~C.} \bibnamefont{Lee}} \bibnamefont{and}
  \bibinfo{author}{\bibfnamefont{S.~J.} \bibnamefont{Muller}},
  \bibinfo{journal}{Polymer} \textbf{\bibinfo{volume}{40}},
  \bibinfo{pages}{2501} (\bibinfo{year}{1999}).

\bibitem[{\citenamefont{Pierleoni and Ryckaert}(1995)}]{pierryck95}
\bibinfo{author}{\bibfnamefont{C.}~\bibnamefont{Pierleoni}} \bibnamefont{and}
  \bibinfo{author}{\bibfnamefont{J.-P.} \bibnamefont{Ryckaert}},
  \bibinfo{journal}{Macromolecules} \textbf{\bibinfo{volume}{28}},
  \bibinfo{pages}{5097} (\bibinfo{year}{1995}).

\bibitem[{\citenamefont{Link and Springer}(1993)}]{linspr93}
\bibinfo{author}{\bibfnamefont{A.}~\bibnamefont{Link}} \bibnamefont{and}
  \bibinfo{author}{\bibfnamefont{J.}~\bibnamefont{Springer}},
  \bibinfo{journal}{Macromeolecules} \textbf{\bibinfo{volume}{26}},
  \bibinfo{pages}{464} (\bibinfo{year}{1993}).

\bibitem[{\citenamefont{Bossart and {\"{O}}ttinger}(1997)}]{bossott97}
\bibinfo{author}{\bibfnamefont{J.}~\bibnamefont{Bossart}} \bibnamefont{and}
  \bibinfo{author}{\bibfnamefont{H.~C.} \bibnamefont{{\"{O}}ttinger}},
  \bibinfo{journal}{Macromolecules} \textbf{\bibinfo{volume}{30}},
  \bibinfo{pages}{5527} (\bibinfo{year}{1997}).

\end{thebibliography}

\end{document}